\documentclass[aps, twocolumn, amsmath, amssymb, superscriptaddress, nofootinbib]{revtex4-1}

\usepackage[utf8]{inputenc}
\usepackage{graphicx}
\usepackage{units}
\usepackage{color}
\usepackage[pdftex, colorlinks=true, linkcolor=myblue, citecolor=myblue, urlcolor=myblue]{hyperref}
\usepackage{times}
\usepackage{comment}
\usepackage{graphicx}
\usepackage{amsmath, calc}
\usepackage{mathrsfs}
\usepackage{amsfonts}
\usepackage{amssymb}
\usepackage{bm}
\usepackage{bbm}
\usepackage{color}
\usepackage{array}
\usepackage{units}

\definecolor{myblue}{rgb}{0,0,1}
\definecolor{myred}{rgb}{1,0,0}

\begin{document}

\title{Dipolar and Quadrupolar Excitons Coupled to a Nanoparticle-on-a-Mirror Cavity}

\author{A. Cuartero-Gonz\'alez}
\affiliation{Departamento de F\'isica Te\'orica de la Materia
Condensada and Condensed Matter Physics Center (IFIMAC),
Universidad Aut\'onoma de Madrid, E-28049 Madrid, Spain}

\author{A. I. Fern\'{a}ndez-Dom\'{i}nguez}
\email{a.fernandez-dominguez@uam.es} \affiliation{Departamento de
F\'isica Te\'orica de la Materia Condensada and Condensed Matter
Physics Center (IFIMAC), Universidad Aut\'onoma de Madrid, E-28049
Madrid, Spain}

\begin{abstract}
We investigate plasmon-emitter interactions in a
nanoparticle-on-a-mirror cavity. We consider two different sorts
of emitters, those that sustain dipolar transitions, and those
hosting only quadrupolar, dipole-inactive, excitons. By means of a
fully analytical two-dimensional transformation optics approach,
we calculate the light-matter coupling strengths for the full
plasmonic spectrum supported by the nanocavity. We reveal the
impact of finite-size effects in the exciton charge distribution
and describe the population dynamics in a spontaneous emission
configuration. Pushing our model beyond the quasi-static
approximation, we extract the plasmonic dipole moments, which
enables us to calculate the far-field scattering spectrum of the
hybrid plasmon-emitter system. Our findings, tested against fully
numerical simulations, reveal the similarities and differences
between the strong coupling phenomenology for bright and dark
excitons in nanocavities.
\end{abstract}
\maketitle

\section{Introduction} \label{sec:intro}

Plasmonic nanostructures allow tailoring the emission
characteristics of microscopic light sources~\cite{Giannini2011}.
The ability of surface plasmons (SPs) to modify the local density
of photonic states and the near-to-far-field coupling efficiency
of nanoemitters was firstly exploited for the conception, design
and optimization of optical
nanoantennas~\cite{Bharadwaj2009,Biagioni2012}. In this context
,
the goal was amplifying both physical quantities to improve the
inherent radiative properties of dye molecules and quantum dots,
mainly. These nanoantenna-enhanced (faster, brighter or
directional) emitters have found applications in areas such as
photodetection, nonlinear optics or imaging~\cite{Novotny2011}.

More recently, the scientific and technological focus in
nanophotonics has shifted from the classical to the quantum
optical regime~\cite{Chang2006,Tame2013}. In the quest for
ultra-compact nonclassical light sources, new design strategies
for plasmonic devices are required~\cite{Fernandez-Dominguez2018}.
In order to develop functionalities operating with single
photons~\cite{Hartsfield2015,Gong2015,Hoang2016}, the interaction
between quantum emitters (QEs) and the electromagnetic (EM) fields
associated with SPs must be enhanced, and even pushed beyond the
weak coupling regime~\cite{Hummer2013}. This demands
nanostructures yielding extremely large and highly structured
spectral densities, in a similar way as nanoantennas do. On the
contrary, in order to minimize radiation losses, which constraints
the QE-SP interaction strength, the near-field of QEs must be
effectively decoupled from their far-field. For this reason,
plasmonic resonators perform as nanocavities for quantum optical
applications~\cite{Marquier2017}. Importantly, despite of their
low quality factor (caused by metal absorption), the small
effective volume of SPs give access to an unexplored parametric
region of light-matter coupling, not accessible by other photonic
technologies~\cite{Schwartz2011,Zengin2015,DeGiorgi2018}.

In the strong coupling regime, light and matter excitations mix
together, giving rise to hybrid states known as plasmon-exciton
polaritons (PEPs). PEP characteristics can be controlled through
the weight of their two constituents~\cite{Chikkaraddy2016}. For
quantum nanophotonics applications, this phenomenon makes it
possible to tune the balance between the high coherence of SPs,
and the high nonlinearities of optical transitions in
QEs~\cite{Torma2015}. Additionally, the small mode volume of SPs
open the way to the formation of PEPs at the single QE level at
room temperature. Thus, much research efforts have concentrated in
this objective in the last years. Experimental evidence of strong
coupling in ensembles of very few, or even single, QEs has been
reported in various gap nanocavities~\cite{Baumberg2019}:
nanoparticle dimers~\cite{Santhosh2016,Roller2016}, tip-substrate
nanoresonators~\cite{Gross2018,Park2019} or nanoparticle-on-mirror
(NPoM) configurations~\cite{Chikkaraddy2016,Kleeman2017,Leng2018}.
Similarly, theoretical advances have revealed the geometric and
material conditions most convenient for the realization of
plasmonic strong coupling at the single emitter
level~\cite{Savasta2010,Manjavacas2011,Delga2014,Gonzalez-Tudela2014,Li2016},
as well as strategies to harness photon correlations in hybrid
QE-SP systems~\cite{Ridolfo2010,Gonzalez-Tudela2013,
Alpeggiani2016,Saez-Blazquez2017}.

Apart from the strong light-matter interactions that they enable,
SPs bring other opportunities to the emerging field of quantum
nano-optics. Their deeply sub-wavelength nature unlocks attributes
of QEs that are elusive to propagating EM fields, such as chiral
phenomena associated to their
polarization~\cite{Lodahl2017,Downing2019}, mesoscopic
effects~\cite{Andersen2010,Neuman2018}, individual atomic
bonds~\cite{Benz2016,Carnegie2018} or light-forbidden
excitons~\cite{Zurita2002,Rivera2017}. Among the latter, special
attention has been paid to quadrupolar excitons, as there are
several theoretical
predictions~\cite{Filter2012,Yannopapas2015,Sanders2018,Cuartero-Gonzalez2018}
that suggest that these can be brought up to time scales
comparable to dipolar ones, even in the strong coupling regime.

In this Article, we investigate the interaction between the SPs
supported by a NPoM cavity and single QEs of two types: those
sustaining dipolar, and those supporting quadrupolar excitons. We
present a transformation optics~\cite{Pendry2012,Luo2013} (TO)
approach which allows us to obtain analytical expressions for all
the physical magnitudes characterizing the hybrid QE-SP system. In
this context, TO allows the quantization of the plasmonic modes
supported by the NPoM cavity and the parametrization of the QE-SP
interaction Hamiltonian, in a similar way as other recently
proposed methods~\cite{Dzsotjan2016,Hughes2018,Franke2019}. Our
approach sheds deep insights into two different plasmon-exciton
phenomena: the near-field population dynamics in a spontaneous
decay configuration and the far-field scattering spectra under
dark-field laser illumination. Our model also accounts for
finite-size effects associated with both excitonic charge
distributions. Throughout the text, the differences and
similarities in the light-matter strong coupling phenomenology for
dipolar and quadrupolar transitions are discussed and analyzed.
Despite the fact that a full three-dimensional (3D) TO framework
for dipolar point-like sources is available~\cite{Li2016}, we
employ here its two-dimensional (2D) version~\cite{Aubry2011}.
There are three reasons justifying this choice. First, only the
latter is fully analytical, which is instrumental for the
description of quadrupolar
transitions~\cite{Cuartero-Gonzalez2018} and finite-sized QEs.
Second, the 2D theory can be pushed beyond purely quasi-static
approximation~\cite{Demetriadou2017}, which allows computing the
SP dipole moments and therefore, far-field spectra for the QE-SP
system. Finally, the comparison between 2D and 3D results reveals
that the 2D treatment reproduces all the phenomenology reported in
3D~\cite{Li2018}.

The paper is organized as follows: Section~\ref{sec:TO} outlines
the fundamental aspects of our TO approach. The spectral densities
and SP coupling strengths for dipolar and quadrupolar QEs are
presented in Sec.~\ref{sec:density}. Section~\ref{sec:finite}
describes mesoscopic effects associated with the finite size of
excitonic charge distributions. In Sec.~\ref{sec:dynamics}, we
study the dynamics of the population exchange between QE and SPs
in an spontaneous emission configuration. The calculation of the
SP dipole moments and the scattering spectrum of the nanocavity
under coherent pumping is described in Section~\ref{sec:spectrum}.
Finally, general conclusions are raised in
Sec.~\ref{sec:conclusions}

\section{Transformation Optics approach} \label{sec:TO}

\begin{figure}[!t]
\includegraphics[width=1\linewidth]{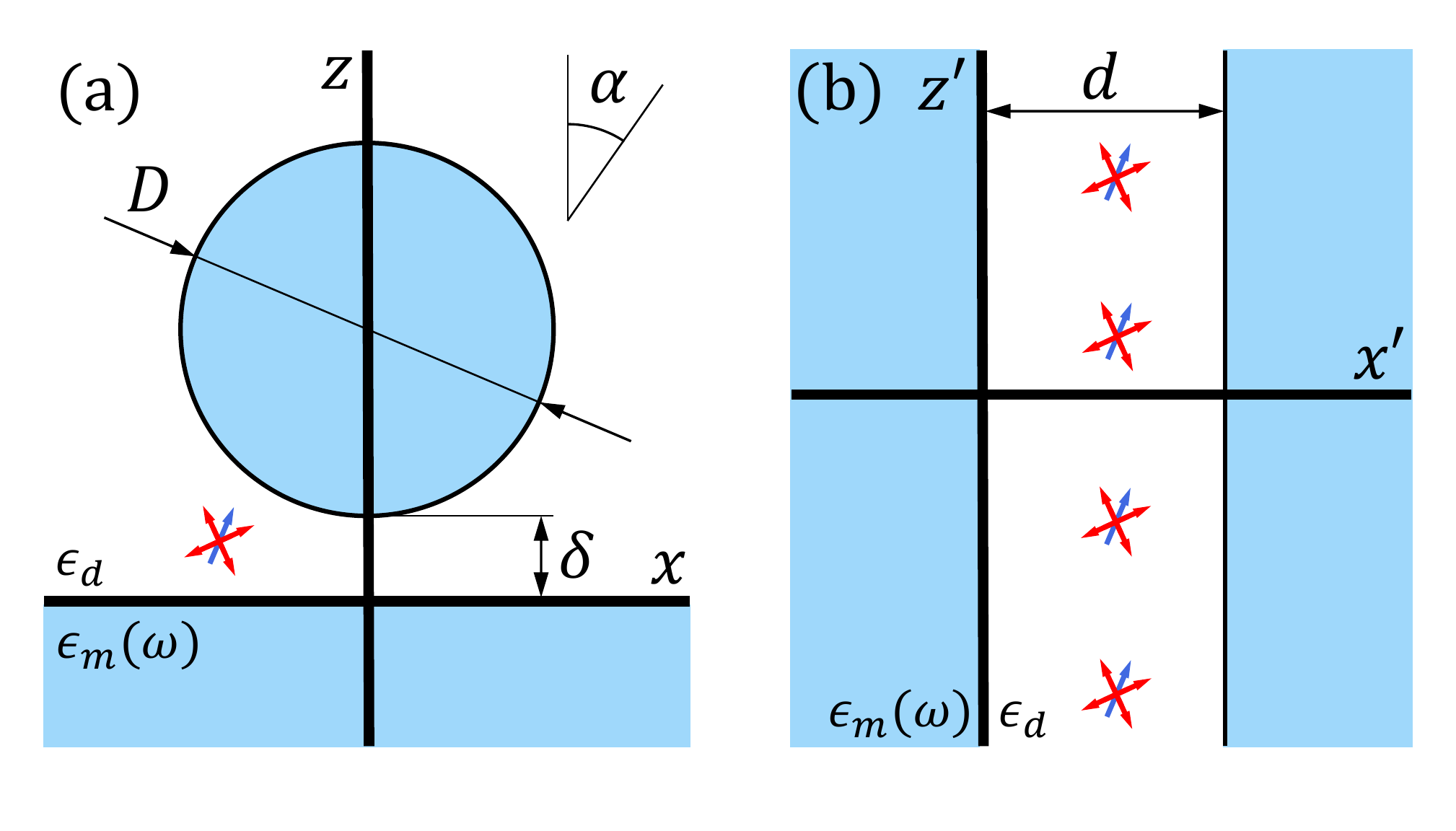}
\vspace{-0.35 cm} \caption{Sketch of the QE-SP system. (a) A
microscopic dipolar (blue) or quadrupolar (red) light source
coupled to the SPs supported by a NPoM cavity with gap $\delta$
and diameter $D$. (b) Geometry obtained under the mapping given by
Equation~\eqref{eq:map}: a metal-dielectric-metal structure
excited by a periodic array of transformed dipole and quadrupole
sources. Note the unprimed and primed coordinates used for
original and transformed geometries.} \label{fig:1}
\end{figure}

The system under study is depicted in Figure~\ref{fig:1}(a): the
interaction of a single QE with a NPoM plasmonic cavity
characterized by the gap $\delta$ and nanoparticle diameter $D$
(related by the ratio $\rho = \delta/D$). The metal permittivity
is given by a Drude fitting to silver,
$\epsilon_m(\omega)=\epsilon_{\infty}-\omega_p^2/(\omega(\omega+i\gamma_m))$
with $\epsilon_{\infty} = 9.7$, $\omega_p = 8.91$ eV and
$\gamma_m= 0.06$ eV. The cavity is embedded in a dielectric medium
with permittivity $\epsilon_d = 4$, which models a DNA origami
scaffolding \cite{Kaminska2018}. The QE, which can be located
anywhere in the surroundings of the nanostructure, is modelled as
a point-like EM source of dipolar or quadrupolar character. It is
parameterized by its dipole, $\boldsymbol{\mu}$ or quadrupole,
$\textbf{Q}$, moment. In order to describe light-matter
interactions in this system, we benefit from the TO formalism
previously developed to compute the absorption and scattering
cross section of similar systems~\cite{Aubry2011}. We consider the
2D version of the geometry in Figure~\ref{fig:1}(a), which has
translational invariance along the out-of-plane direction. This
simplifies greatly the calculation of the EM fields scattered by
the NPoM geometry under point dipole and quadrupole excitations,
which can be tackled fully analytically.

Let us introduce first the description of the QE sources in the 2D
model. The dipolar exciton is characterized by its vectorial
dipole moment $\boldsymbol{\mu}=(\mu_x,
\mu_z)=\mu(\sin\alpha,\cos\alpha)$, where angle $\alpha$ (see
Fig.~\ref{fig:1}(a)). The quadrupolar exciton is defined by its
tensorial quadrupole moment
\begin{equation}
\textbf{Q} = \left( \begin{array}{cc} Q_{xx} & Q_{xz} \\Q_{zx} &
Q_{zz}  \end{array} \right)= \frac{Q}{\sqrt{2}} \left(
\begin{array}{cc} \sin 2\alpha & \cos 2\alpha \\\cos 2\alpha &
- \sin 2\alpha  \end{array} \right).\label{eq:Q}
\end{equation}
Gauge invariance allows writing the quadrupolar tensor above in a
traceless form, without modifying the light-matter coupling
Hamiltonian~\cite{Novotnybook}. Therefore, by symmetry, there are
only two independent entries, $Q_{xx}$ and $Q_{xz}$ in the tensor,
which can be expressed in terms of the modulus
$Q=\sqrt{\sum_{i,j}Q^2_{ij}}$ and the angle $\alpha$. The
quadrupolar source can be interpreted as composed by 4 point-like
charges (each of them of opposite sign to the adjacent ones)
placed in the vertices of a square. For $\alpha=0$
($\alpha=\pi/4$) the edges (diagonals) of the quadrupolar charge
distribution are parallel to the $x$ and $z$-axes (see
Sec.~\ref{sec:finite}).

We employ the 2D model to obtain the Purcell factor, $P(\omega)$,
experienced by each QE, which we will subsequently use to extract
the corresponding spectral density and QE-SP coupling strengths.
Taking advantage of the nanometric dimensions of the cavity, we
can operate in the quasi-static regime to calculate the Purcell
spectra for both QEs. These are given by the ratio of the power
dissipated by the point-like EM source in the presence of the NPoM
cavity and the embedding medium over the power radiated in free
space
\begin{eqnarray}
P_\mu(\omega)&=&\frac{\omega}{2\,W_\mu^{(0)}(\omega)} \Big|{\rm
Im}\{ \boldsymbol{\mu} \nabla\ \phi_\mu(\textbf{
r},\omega)|_{\textbf{r}_{E}}\} \Big|, \label{eq:Pmu}
\\
P_Q(\omega)&=&\frac{\omega}{2\,W_Q^{(0)}(\omega)} \Big|{\rm
Im}\{(\textbf{Q}\nabla)\nabla\phi_Q(\textbf{r},\omega)|_{\textbf{
r}_{E}}\}  \Big|,\label{eq:PQ}
\end{eqnarray}
where $\textbf{r}=(x,z)$ and $\textbf{ r}_{E}$ is the position of
the emitter in the $xz$-plane.

The radiated powers in the denominator of
Eqs.~\eqref{eq:Pmu}~and~\eqref{eq:PQ} can be computed through the
free-space EM Dyadic Green's function in 2D,
\begin{eqnarray}
W_\mu^{(0)}(\omega)&=&\frac{\omega^3}{2\epsilon_0
c^2}\rm{Im}\{\boldsymbol{\mu}
    \textbf{G}^{(0)}(\textbf{r},\textbf{r}_{E},\omega)|_{\textbf{r}_{E}} \boldsymbol{\mu}\}, \label{eq:mu_free}\\
W_{Q}^{(0)}(\omega)&=&\frac{\omega^3}{2\epsilon_0
c^2}\rm{Im}\{(\textbf{Q}\nabla)
    (\nabla^{'} \textbf{G}^{(0)}(\textbf{r},\textbf{r}',\omega)|_{\textbf{r}_{E}})
    \textbf{Q}\},\label{eq:Q_free}
\end{eqnarray}
where
$\textbf{G}^{(0)}(\textbf{r},\textbf{r}',\omega)=\tfrac{1}{4i}\big[\textbf{I}+\left(\tfrac{\omega}{c}\right)^2\nabla
\nabla\big]H_0^{(1)}(\omega|\textbf{r}-\textbf{r}'|/c)$ and
$H_0^{(1)}(\cdot)$ is the zero-order Hankel function of the First
Kind. The expressions above yield
$W_\mu^{(0)}(\omega)=\mu^2\omega^3/(16\epsilon_0c^2)$ and
$W_{Q}^{(0)}(\omega)=Q^2\omega^5/(64\epsilon_0c^4)$. Note that in
Eqs.~\eqref{eq:Pmu}-\eqref{eq:Q_free} we have used that the power
dissipated by dipole and quadrupole emitters is given by
$W_\mu=\tfrac{\omega}{2}{\rm Im}\{\boldsymbol{\mu}\textbf{E}\}$
and $W_\mu=\tfrac{\omega}{2}{\rm
Im}\{(\textbf{Q}\nabla)\textbf{E}\}$, respectively, where
$\textbf{E}$ stands for the electric field due the source in each
case~\cite{Novotnybook}.

Equations~\eqref{eq:Pmu}~and~\eqref{eq:PQ} show that, neglecting
radiative losses, we only need the quasi-static potentials
$\phi_\mu(\textbf{r},\omega)$ and $\phi_Q(\textbf{r},\omega)$ in
order to determine the Purcell spectra. This we do using TO, by
applying the conformal map
\begin{equation}
\varrho' = \ln\bigg(\frac{ 2iD\sqrt{\rho(1+\rho)}}{\varrho-is} +
1\bigg), \label{eq:map}
\end{equation}
where $s=\delta+D\sqrt{\rho}/(\sqrt{1+\rho}+\sqrt{\rho})$, and
$\varrho' = x' + i z'$ and $\varrho = x + i z$ denote,
respectively, the spatial coordinates (in complex notation) for
the original and transformed frames. Under Eq.~\eqref{eq:map}, the
cavity maps into a metal-dielectric-metal waveguide of width $d =
2\ln(\sqrt{\rho} + \sqrt{1+\rho})$, see Figure~\ref{fig:1}(b).
Importantly the permittivities are not affected by the mapping.
The QE sources transform into a periodic array of sources of the
same character as the initial ones. They are distributed along
$z'$-direction with period $2\pi$. Thus, the Purcell spectra
calculation reduces to solving Laplace's equation in the geometry
of Fig.~\ref{fig:1}(b). The solution in the original frame is
easily obtained using $
\phi_{\mu,Q}(\varrho,\omega)=\phi'_{\mu,Q}(\varrho'(\varrho),\omega)$.
Details of the calculation as well as full analytical expressions
for the quasi-static potentials are given in
Appendix~\ref{apen:potential}.

\begin{figure}[!t]
\includegraphics[width=1\linewidth]{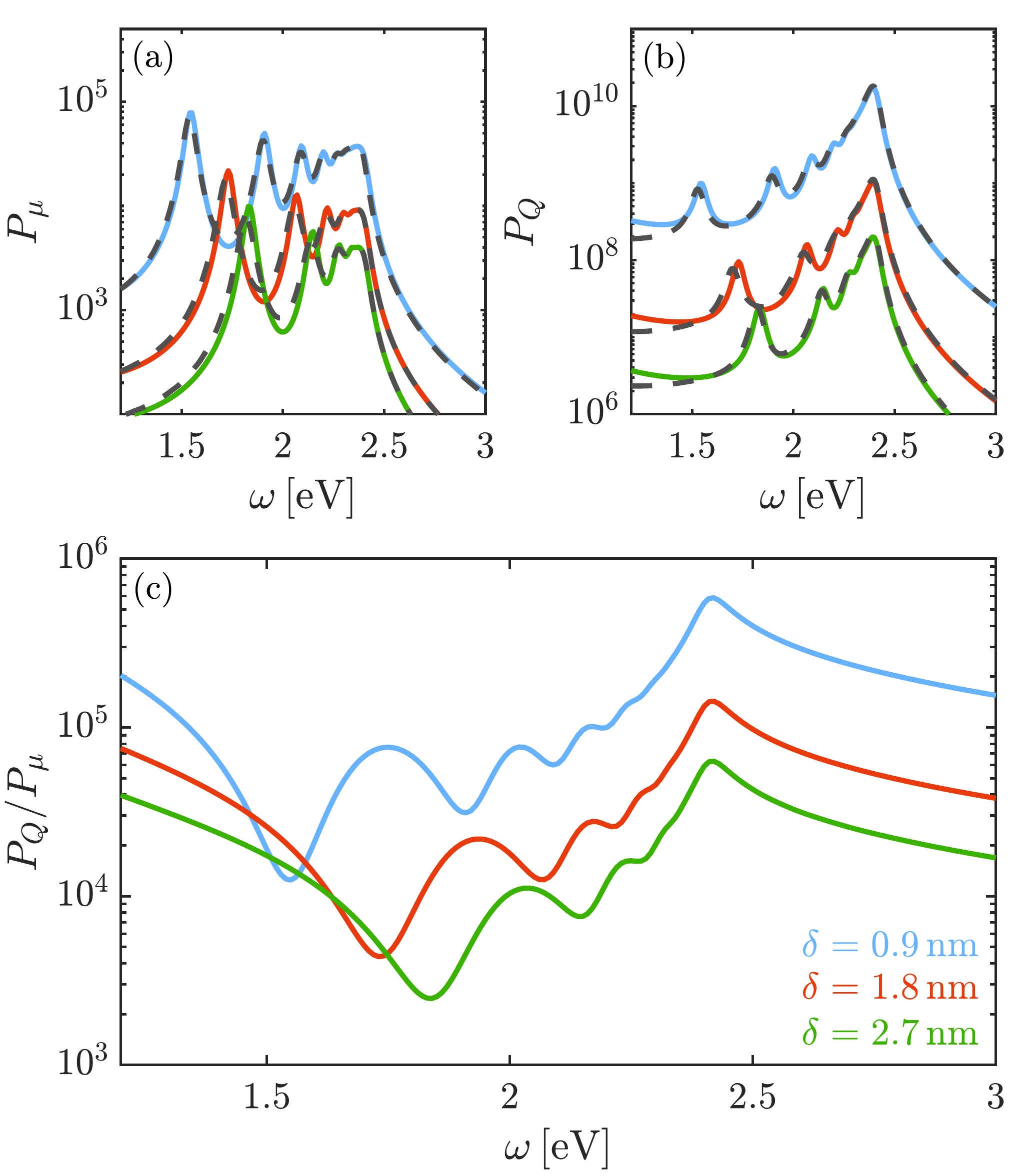}
\vspace{-0.5 cm}\caption{Purcell factor at gap center
($z_{E}=0.5\delta$) of NPoM cavities with $D=30$ nm and three
different gap sizes, $\delta$. (a) Dipolar QE oriented along
$z$-direction. (b) Quadrupolar QE with vanishing diagonal terms in
${\bf Q}$. In both panels, analytical (solid lines) and numerical
(dashed lines) spectra are compared. (c) Ratio between the
quadrupolar and dipolar Purcell factor,
$P_{Q}(\omega)/P_{\mu}(\omega)$ shown above.} \label{fig:2}
\end{figure}

Figure~\ref{fig:2} shows the Purcell spectra for NPoM cavities of
30 nm diameter and different gap sizes: 0.9 nm (blue), 1.8 nm
(red) and 2.7 nm (green). Two different QEs, placed at the gap
center, $z_{E}=0.5\delta$, are considered: (a) a dipolar exciton
oriented along $z$-direction ($\mu_z=\mu$) and (b) a quadrupolar
exciton with a purely non-diagonal moment
($Q_{xz}=Q_{zx}=Q/\sqrt{2}$). In our parametrization, $\alpha=0$
for both QEs. Analytical predictions obtained from our TO approach
(color solid lines) are compared against fully numerical 2D
calculations (black dashed lines) using the Laplace's Equation
solver implemented in Comsol Multiphysics$^{\rm TM}$. Both set of
quasi-static spectra are in very good agreement. There are small
discrepancies in the low frequency tail of Fig.~\ref{fig:2}(b),
which we attribute to the failure of the assumption that the
photonic environment is fully governed by the SP modes supported
by the NPoM cavity. Note that our 2D model slightly underestimates
Purcell factors when compared to full 3D calculations obtained for
the same geometric parameters.

In Fig.~\ref{fig:2}(c), we plot the ratio between the quadrupolar
and dipolar Purcell factors for the three cavity configurations in
the panels above. We observe that $P_{Q}(\omega)$ is several
orders of magnitude larger than $P_{\mu}(\omega)$ throughout the
spectral window considered. However, the ratio between both
magnitudes is largest in two regions, the low frequency tail and
around $\omega_{\rm PS}\simeq2.4$ eV. We anticipate here that the
nanocavities do not support SPs in the former, and that the latter
corresponds to the plasmonic pseudomode formed due to the spectral
overlapping of high-order dark SP modes (see
Refs.~\cite{Delga2014,Li2016}). In this spectral position, which
lies in the vicinity of the surface plasmon asymptotic frequency
(given by the condition $\epsilon_m(\omega)+1=0$~\cite{Li2018}),
the reduction of the gap size enhances quadrupolar transitions the
most, yielding $P_{Q}/P_\mu>10^5$ for $\delta=0.9$ nm. In these
conditions, the time scales of quadrupole and dipole exciton
dynamics become similar, as their decay rates in free-space differ
in 5-6 orders of magnitude~\cite{Rivera2017}.

\section{Spectral Density and Coupling Strengths} \label{sec:density}

The spectral density $J(\omega)$ contains information about the electromagnetic modes supported by the cavity as
well as the coupling strength between each of them and the QE.
Thus, it depends on the cavity geometry (diameter and gap size)
and permittivity, the exciton characteristics (natural frequency
and dipolar/quadrupolar moment) and its position and orientation.
It can be expressed in terms of the Purcell factor as
\begin{equation}
J_i(\omega)=\frac{\gamma_i}{2\pi}P_i(\omega),
\end{equation}
where $i=\mu,Q$ and $\gamma_i$ is the QE decay rate in free-space.
Note that the equation above ensures that, in the weak-coupling
regime, the general Wigner-Weisskopf theory for spontaneous
emission (valid beyond the Markovian approximation) recovers an
exciton population decaying monotonically in time with Purcell
enhanced decay rate $P_i\gamma_i$~\cite{Petruccionebook}. In order
to calculate the spectral density experienced by dipolar and
quadrupolar excitons in the vicinity of the NPoM cavity, we use
the 2D Purcell spectra obtained in the previous section. The decay
rates in free-space are taken from 3D
calculations~\cite{Klimov2005},
$\gamma_\mu=\omega^3\mu^2/(3\pi\epsilon_0\hbar c^3)$ and
$\gamma_Q=\omega^5 Q^2/(360\pi\epsilon_0\hbar c^5)$.

By simple inspection of
Equations~\eqref{eq:pot_mu_sc}~and~\eqref{eq:pot_mu_sc}, we can
see that the Purcell spectra and,  therefore the spectral density
for both dipolar and quadrupolar excitons are composed by a number
of SP contributions (labelled with index $n$, that is related to
the azimuthal order of the plasmonic mode) whose resonant
condition reads
\begin{equation}
(\sqrt{\rho} + \sqrt{1+\rho})^{2n} = \bigg|\text{Re}
\bigg(\frac{\epsilon_{m}(\omega)-\epsilon_d}{\epsilon_{m}(\omega)+\epsilon_d}\bigg)\bigg|,
\label{eq:denom}
\end{equation}
which reproduces the quasi-static condition for absorption maxima
obtained under plane-wave illumination~\cite{Aubry2011}. Note that
Eq.~\eqref{eq:denom} is quadratic, which means that we can
identify two different solutions for a given $n$. These correspond
to SP modes with different parity (with respect to the NPoM gap
cavity), which we label as $\sigma=1$ (even) and $\sigma=-1$
(odd). Note that the term even (odd) refers to the symmetric
(antisymmetric) character of the SP electric fields across the
gap.

Exploiting the Drude form of the metal permittivity, and using the
high quality resonator limit~\cite{Waks2010}, we can expand the
spectral densities as a sum of Lorentzian SP terms of the form
\begin{equation}
J_{i}(\omega)
    = \sum_{n=1}^{\infty} \sum_{\sigma = \pm 1}
    \frac{(g_{i}^{n,\sigma})^2}{\pi}\frac{\gamma_m/2}{(\omega-\omega_{n,\sigma})^2+(\gamma_m/2)^2}, \label{eq:lorentzians}
\end{equation}
where $\gamma_m$ is the Drude damping rate (note that SP radiative
damping is neglected in the quasi-static limit) and
\begin{equation}
\omega_{n,\sigma} = \frac{\omega_p}{\sqrt{\epsilon_{\infty} +
\epsilon_d\xi_{n,\sigma}}} \label{eq:SPfreqs}
\end{equation}
are the SP frequencies, with $\xi_{n,\sigma} = \frac{(\sqrt{\rho}
+ \sqrt{1 + \rho})^{2n} + \sigma}{(\sqrt{\rho} + \sqrt{1 +
\rho})^{2n} - \sigma}$. Observe the remarkable similarity of the
expression above with its 3D counterpart~\cite{Li2018}.

Equation~\eqref{eq:SPfreqs} shows that for both parities, SPs with
increasing $n$ approach the pseudomode frequency, $\omega_{\rm
SP}=\frac{\omega_p}{\sqrt{\epsilon_{\infty} + \epsilon_d}}$. Large
$\rho$ provides faster convergence of the SPs frequencies to
$\omega_{\rm SP}$. It also reveals that the frequency of even
($\sigma=+1$) modes increase towards this value, whereas it
decreases for the odd ($\sigma=-1$) ones. As expected from the
transformed geometry in Fig.~\ref{fig:1}, this phenomenology is
equivalent to the dispersion of the SPs supported by
metal-dielectric-metal waveguides, where both even and odd bands
approach asymptotically the frequency for the single
metal-dielectric interface (given by $\omega_{\rm PS}$ above).

The weight of each term in Equation~\eqref{eq:lorentzians} is
given by $g_{i}^{n,\sigma}$, the QE-SP coupling strength between
the dipolar or quadrupolar exciton ($i=\mu,Q$) and the SP mode of
azimuthal order $n$ and parity $\sigma$. These constants contain
all the information about the QEs and the SP mode spatial profile.
Appendix~\ref{apen:gs} presents the analytical expressions for
$g_{i}^{n,\sigma}$ that we obtain from our TO approach. Their
dependence on the QE position and orientation is analyzed in
detail below.

\begin{figure}[!t]
\includegraphics[width=0.95\linewidth]{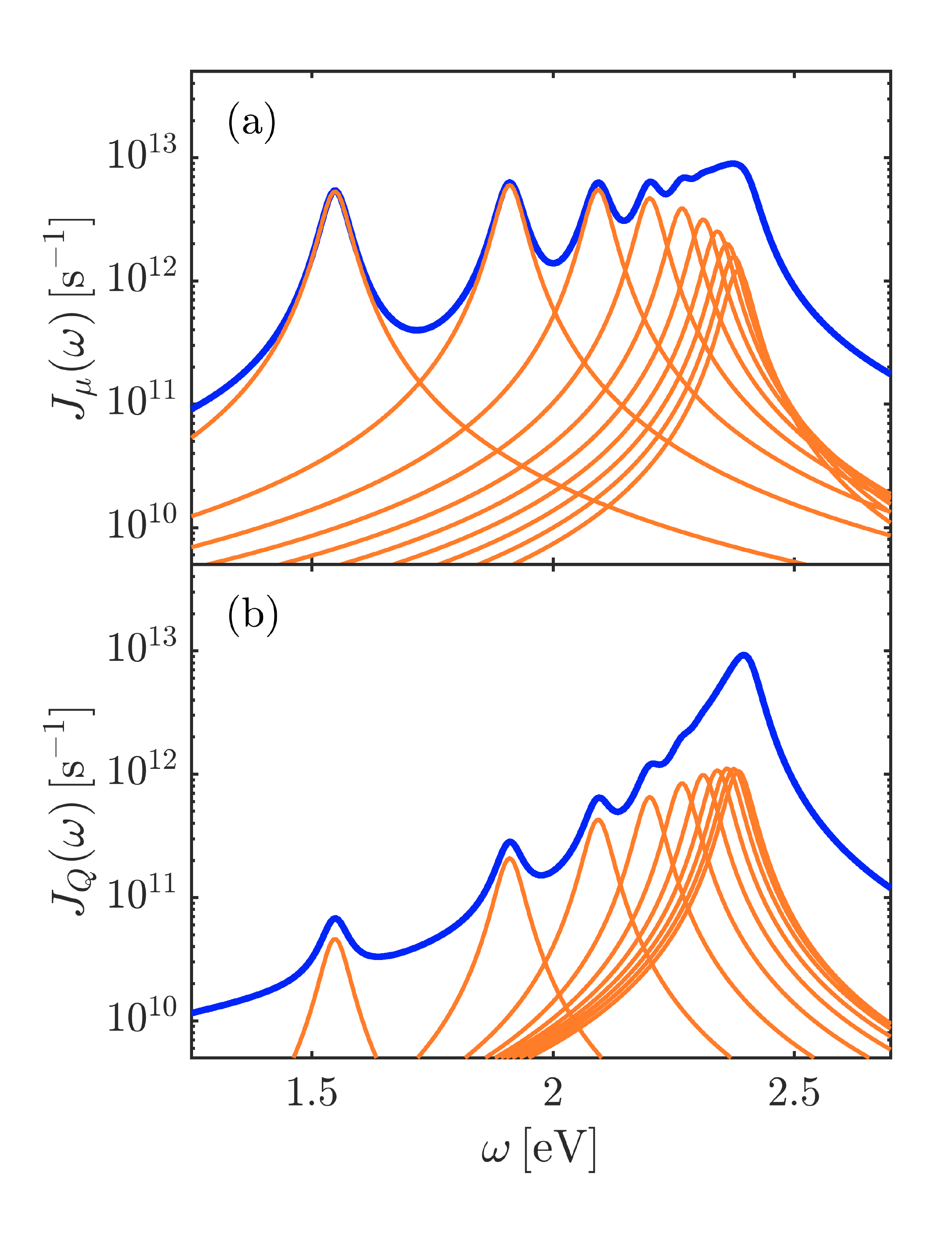}
\vspace{-0.15 cm}\caption{Spectral density (blue line) at the gap
center $(x_E,z_E)=(0,0.5\delta)$ of the NPoM cavity with $D = 30$
nm and $\delta = 0.9$ nm: (a) dipolar QE with $\mu=0.55$
{e$\cdot$nm} and (b) quadrupolar QE with $Q=0.75$
{e$\cdot$nm$^2$}. The orientation in both cases is the same as in
Fig.~\ref{fig:2} ($\alpha=0$). Orange lines plot the first few
terms ($n\leq10$) in the decomposition in
Eq.~\eqref{eq:lorentzians}. By symmetry, the QEs are coupled only
to SP modes with $\sigma=1$.} \label{fig:3}
\end{figure}

Figure~\ref{fig:3} shows the spectral densities for a dipolar (a)
and a quadrupolar (b) exciton in the same cavity configuration as
the blue line in Figure~\ref{fig:2}. The QEs are placed at the gap
center and their orientation is $\alpha=0$. In agreement with
experimental values, we set $\mu=0.55$ {e$\cdot$nm}. We take
$Q=0.75$ {e$\cdot$nm$^2$} for the quadrupole moment, which yields
a ratio between free-space decay rates
$\gamma_Q/\gamma_\mu\simeq1\cdot10^{-6}$ at the center of the
frequency window considered, $\omega=2$ eV. Despite this inherent
difference between both QEs, the spectral densities are equivalent
at the pseudomode position, $J_\mu(\omega_{\rm
PS})=J_Q(\omega_{\rm PS})\simeq10^{-13}\,{\rm s}^{-1}$. Therefore,
the interaction strength between the NPoM cavity and both QEs at
$\omega_{\rm PS}\simeq2.4$ eV is very similar. On the contrary, at
lower frequencies, $J_\mu(\omega)\gg J_Q(\omega)$, which indicates
that the quadrupole exciton couples more weakly than the dipolar
one to low-order SPs (with small $n$). Note the large contrast at
the dipolar SP,
$J_\mu(\omega_{1,+1})\simeq10^2J_Q(\omega\omega_{1,+1})$. In fact,
the maximum at $\omega_{1,+1}=1.55$ eV barely stands out of the
low-frequency tail of the pseudomode maximum in the quadrupolar
spectral density, see Fig.~\ref{fig:3}(b).

\begin{figure}[!t]
\includegraphics[width=1\linewidth]{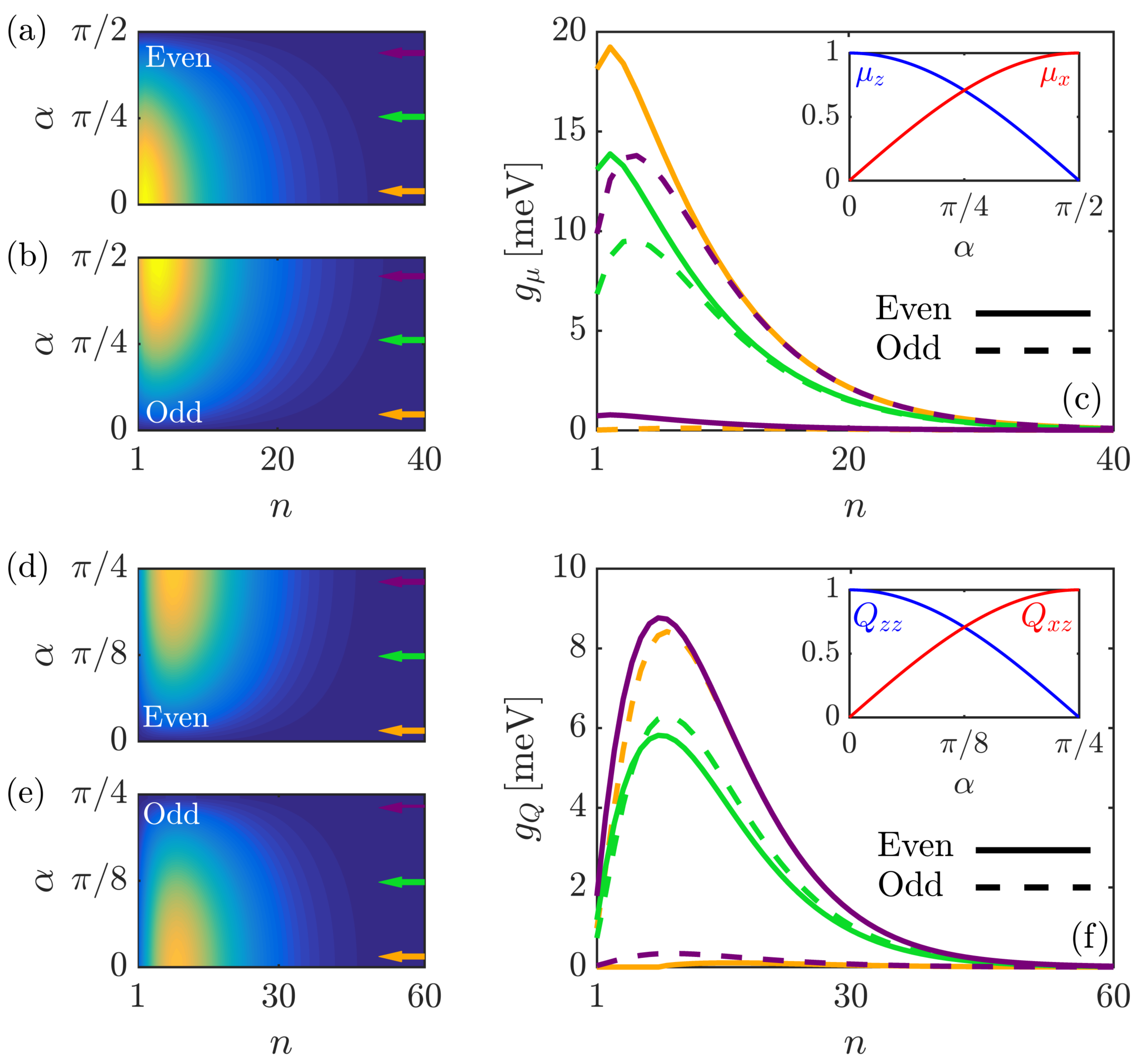}
\vspace{-0.45 cm}\caption{Dependence of the coupling strength on
$n$ and $\alpha$ for dipolar (a-c) and quadrupolar (d-f) excitons
at the gap center. Panels (a) and (b) [(d) and (e)] display
$g_\mu$ [$g_Q$] contourplots for $\sigma=+1$ and $\sigma=-1$,
respectively. Panels (c) and (f) show cuts for even (solid lines)
and odd (dashed lines) modes and three different values of
$\alpha$ (indicated by colored horizontal arrows in the
contourplots). The insets render the dipole and quadrupole
components as a function of $\alpha$.} \label{fig:4}
\end{figure}

In Figure~\ref{fig:4}, we study the dependence of the light-matter
coupling strengths on the QE orientation for the different SPs
supported by the NPoM cavity. Both dipolar ($\mu=0.55$
{e$\cdot$nm}) and quadrupolar ($Q=0.75$ {e$\cdot$nm$^2$}) excitons
are placed at the gap center $(x_E,z_E) = (0,0.5\delta)$.
Fig.~\ref{fig:4}(a)~and~(b) display $g_\mu^{n,+1}$ and
$g_\mu^{n,-1}$, respectively, versus $n$ and $\alpha$. These
conturplots show that the maximum coupling takes place at $n<4$
for both $\sigma$, and that vertical (horizontal) dipolar QEs
couple more efficiently to even (odd) SPs. This can be clearly
seen in Fig.~\ref{fig:4}(c) which plot $g_\mu$ for the three
orientations indicated by arrows in the previous panels. Only for
$\alpha=\pi/4$ (green lines), the coupling strength to even
(solid) and odd (dashed) SPs are comparable. Note that
$g_\mu^{n,+1}>g_\mu^{n,-1}$ for very low azimuthal order, $n$, in
this case. Fig.~\ref{fig:4}(d)~and~(e) display coupling strength
maps for $g_Q^{n,+1}$ and $g_Q^{n,-1}$. They exhibit a similar
dependence on $n$ and $\alpha$ as their dipolar counterparts.
However, two main differences can be observed. First, although
$g_\mu\simeq g_Q$ for large $n$, the maximum coupling is always
lower for quadrupolar QEs within the range of geometric and
material parameters considered. Second, the peak in
$g_i^{n,\sigma}$ always takes place at lower $n$ for the dipolar
excitons. These two circumstances are apparent in
Fig.~\ref{fig:4}(f), which also shows that for a given $n$,
$g_Q^{n,+1}\simeq g_Q^{n,-1}$ at $\alpha=\pi/8$ (green lines).

\begin{figure}[!t]
\includegraphics[width=1\linewidth]{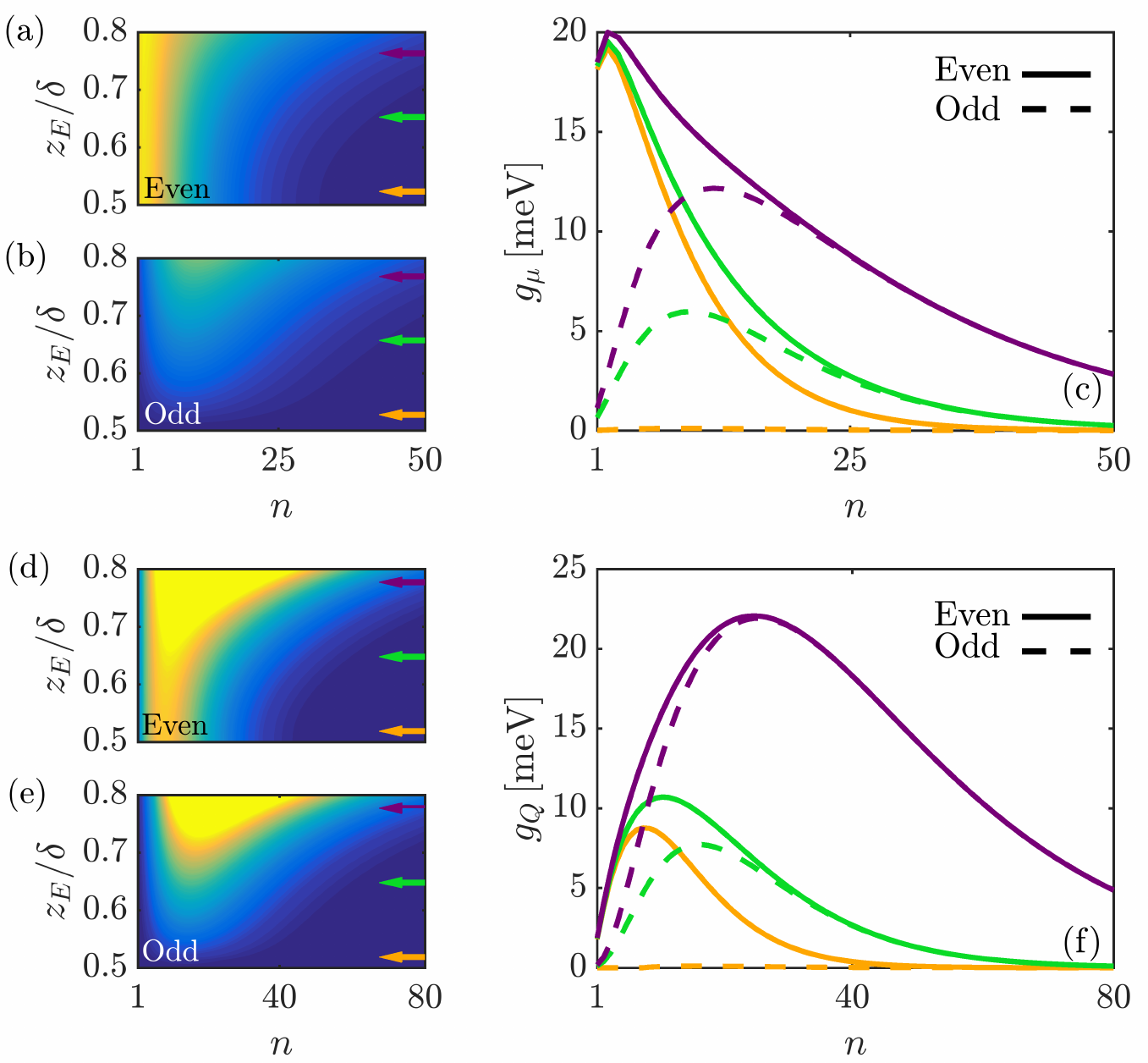}
\vspace{-0.45 cm}\caption{Dependence of the coupling strength on
$n$ and $z_E$ along the symmetry axis of the cavity ($x_E=0$) for
dipolar (a-c) and quadrupolar (d-f) excitons with $\alpha=0$.
Panels (a) and (b) [(d) and (e)] display $g_\mu$ [$g_Q$]
contourplots for $\sigma=+1$ and $\sigma=-1$, respectively. Panels
(c) and (f) show cuts for even (solid lines) and odd (dashed
lines) modes and three different values of of $z_E/\delta$
(indicated by colored horizontal arrows in the contourplots).}
\label{fig:5}
\end{figure}

Once we have studied the orientation-dependence of QE-SP
couplings, we investigate next the impact of the emitter position.
We restrict our attention first to the symmetry axis of the cavity
($x_E=0$). Figure~\ref{fig:5}(a)-(b) display $g_\mu$ maps as a
function of $n$ and $z_E/\delta$ between the gap center and the
vicinity of the NP surface for both plasmonic parities. We can see
that, in accordance with Fig.~\ref{fig:4}, the light-matter
interaction is governed by low-order ($n<4$) even SPs. Note that
the QE coupling to these modes barely depends on the emitter
position. The associated electric field profile is constant along
the NPoM gap. On the contrary, in accordance with the
phenomenology reported for full 3D
models~\cite{Manjavacas2011,Li2016}, $g_\mu$ for both even and odd
modes of higher $n$ increases as the QE approach the NP surface.
This trend is visible in Fig.~\ref{fig:5}(c), which evaluates
$g_\mu^{n,\sigma}$ for three $z_E$ values. A similar analysis is
presented in Fig.~\ref{fig:5}(d)-(f) for quadrupolar excitons. The
$g_Q$ maps reveal that, in contrast to dipolar QEs, the coupling
vanishes for SPs with very low $n$, and the maximum takes place
for $n>10$. As we have already discussed, the QE only interacts
with $\sigma=+1$ modes at the gap center. The emitter displacement
towards the NP surface increases both $g_Q^{n,+1}$ and
$g_Q^{n,-1}$. The light-matter interaction is then fully governed
by the plasmonic pseudomode. Indeed, the cuts at fixed QE
position in Fig.~\ref{fig:5}(f) show that the coupling to even and
odd SPs for large $n$ is maximum, and very similar, at
$z_E=0.75\delta$. Importantly, the maximum coupling in this panel
is higher than in Fig.~\ref{fig:5}(c). This indicates that, by
displacing the emitter from the gap center, the plasmonic
interaction for quadrupole QEs becomes larger than for dipolar
ones.

\begin{figure}[!t]
\includegraphics[width=1\linewidth]{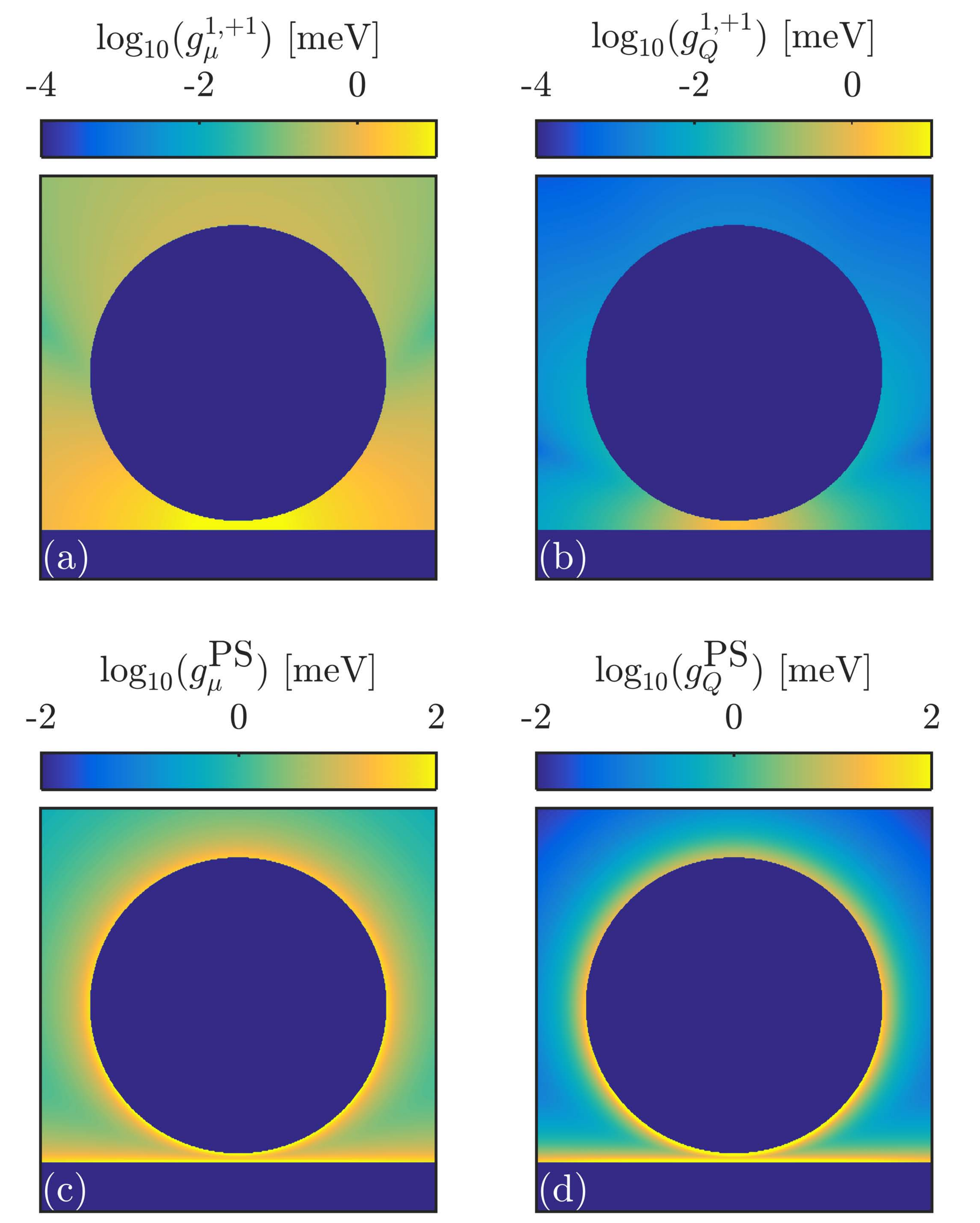}
\vspace{-0.25 cm} \caption{Spatial dependence of QE-SP coupling
strengths (in log scale) within the $xz$-plane ($D=30$ nm,
$\delta=0.9$ nm). Top panels show the (a) dipolar and (b) quadrupolar
exciton coupling strength to the lowest order, even SP mode
($n=1$, $\sigma=+1$). Bottom panels correspond to the (c) dipolar
and (d) quadrupolar coupling to the plasmonic pseudomode.}
\label{fig:6}
\end{figure}

We exploit the analytical power of our TO approach further and
explore fully the spatial distribution of the QE-SP coupling,
$g_i^{n,\sigma}(\varrho_E,\omega)$, in the vicinity of the NPoM
geometry. Figure~\ref{fig:6} displays strength maps (in log scale)
involving the dipolar (a,c) and quadrupolar (b,d) excitons and the
lowest-frequency SP $(\omega_{1,+1})$ (a,b) and the plasmonic pseudomode $(\omega_{\rm PS})$ (c,d). We
set all the parameters as in Fig.~\ref{fig:5}. The former mode
corresponds to $n=1,\sigma = +1$, the coupling constant for the
latter is calculated as~\cite{Delga2014}
\begin{equation}
g_i^{\rm PS}=\sqrt{\sum_{\sigma=\pm1}\sum^\infty_{n=n_{\textrm
{min}}}(g_i^{n,\sigma})^2},\label{eq:gPS}
\end{equation}
where the minimum order for even/odd parity is set by the
condition $|\omega_{\rm
PS}-\omega_{n_{\textrm{min}},\sigma}|\leq\gamma_m/2$. Notice that
$n_{\textrm{min}}=7$ in Fig.~\ref{fig:6}(c)-(d) for both parities,
which is in accordance with Fig.~\ref{fig:3}, which only shows
five distinguishable peaks in $J_i(\omega))$ below $\omega_{\rm
PS}$.

Figure~\ref{fig:6}(a)~and~(b) evidence that the coupling-strength
maps associated to the lowest (dipolar) SP are focused within the
gap of the NPoM geometry. However, the localization at the gap is
significantly larger for the quadrupole QE ($\alpha=0$ for both
excitons). Whereas the region of high $g_\mu$ spreads over the
flat metal surface and the perimeter of the NP, $g_Q$ decays
abruptly within a few nanometer range from the gap center. Let us
remark again that all contourplots are in logarithmic scale. In
contrast, Fig.~\ref{fig:6}(c)~and~(d) demonstrate that the
pseudomode yields coupling maps insensitive to the cavity
geometry. These are much more tightly bounded to the metal
boundaries, within a sub-nm length scale, both at the NP and
substrate surface. The gap does not seem to play any role in the
spatial distribution of $g_\mu$ and $g_Q$, except from the
overlapping of their tails across it. In accordance with the top
panels, the quadrupole distribution is also more confined than the
dipolar one. The remarkable contrast between
Fig.~\ref{fig:6}(a)~and~(d) reveals that through the exploitation
of higher order SP modes and multipolar excitons, spatial
resolutions in the light-matter coupling well below the nanometer
can be achieved~\cite{Benz2016}.

\section{Finite-size Effects} \label{sec:finite}

In this section, we extend our TO approach in order to address the
emergence of mesoscopic effects~\cite{Andersen2010} in the
light-matter interactions due to the finite size of the
QE~\cite{Neuman2018}. The extreme confinement of the plasmonic
coupling strength maps shown in Fig.~\ref{fig:6} suggests that our
NPoM cavity is an ideal platform to explore excitonic charge
distributions beyond the point-like description of the EM source.
As the spatial variation of $g_\mu$ or $g_Q$ approaches
length-scales comparable to the QE dimensions, we can expect that
this approximation breaks down. By inspection of Fig.~\ref{fig:6},
we can anticipate that these finite-size effects are higher for
the plasmonic pseudomode than for SPs with low $n$.

A dipole EM source can be depicted as a pair of point-like
charges of opposite sign and magnitude $|q|$. The vector between
both charge positions is
$\boldsymbol{\ell}=\mu/|q|(\sin\alpha,\cos\alpha)$ (note that we
assume $\mu = |q| \ell $). The Purcell factor will no longer be
given by Eq.~\eqref{eq:Pmu}. Instead, it reads now
\begin{eqnarray}
P_\mu^{\rm ext}(\omega)&=&\frac{\omega\mu/\ell}{2
W_\mu^{(0)}(\omega)}
\Big|\int_{\textbf{r}_E-\tfrac{\boldsymbol{\ell}}{2}}^{\textbf{r}_E+\tfrac{\boldsymbol{\ell}}{2}}
{\rm Im}\big\{\nabla \phi_\ell^{(2)}(\textbf{r}) d\textbf{r}\}\Big|= \notag \\
&=& \frac{\omega\mu/\ell}{2 W_\mu^{(0)}(\omega)} {\rm
Im}\{\phi_\ell^{(2)}\left(\textbf{r}_E-\tfrac{\boldsymbol{\ell}}{2}\right)
-\phi_\ell^{(2)}\left(\textbf{r}_E+\tfrac{\boldsymbol{\ell}}{2}\right)\big\}, \notag \\
\label{eq:Pmuext}
\end{eqnarray}
where, as the QE dimensions are much smaller than optical
wavelengths ($\ell\ll 2\pi c/\omega$), $W_\mu^{(0)}(\omega)$ is
given by Eq.~\eqref{eq:mu_free}. The quasi-static potential
$\phi_\ell^{(2)}(\textbf{r})$ describes the EM fields scattered by
the NPoM cavity excited by the two opposite charges separated a a
distance $\ell$ (note that, for simplicity, we have dropped its
frequency dependence above). The analytical expression for the
potential generated by any neutral distribution of point-like
charges, used to compute $\phi_\ell^{(2)}(\textbf{r})$, is
provided in Appendix~\ref{apen:potential}.

An extended quadrupole source corresponds to a square-shaped
distribution of four point-like charges with side vectors
$\boldsymbol{\ell}=(Q/\sqrt{2}|q|)^{1/2}(\sin\alpha,\cos\alpha)$
and
$\boldsymbol{\ell_\bot}=(Q/\sqrt{2}|q|)^{1/2}(\cos\alpha,-\sin\alpha)$
($Q=\sqrt{2}|q|\ell^2$). The Purcell factor in this case is given
by
\begin{eqnarray}
P_Q^{\rm ext}(\omega)&=&\frac{\omega Q/\ell^2}{2
W_Q^{(0)}(\omega)} {\rm
Im}\Big\{\phi_\ell^{(4)}\big(\textbf{r}_E+\tfrac{\boldsymbol{\ell}_+}{2}\big)
+\phi_\ell^{(4)}\big(\textbf{r}_E-\tfrac{\boldsymbol{\ell}_+}{2}\big)
\notag \\
&&-\phi_\ell^{(4)}\big(\textbf{r}_E+\tfrac{\boldsymbol{\ell}_-}{2}\big)
-\phi_\ell^{(4)}\big(\textbf{r}_E-\tfrac{\boldsymbol{\ell}_-}{2}\big)\Big\},
\label{eq:PQext}
\end{eqnarray}
where
$\boldsymbol{\ell}_\pm=\boldsymbol{\ell}\pm\boldsymbol{\ell_\bot}$,
and $W_Q^{(0)}(\omega)$ is given by Eq.~\eqref{eq:Q_free}. The
analytical expressions used to evaluate
$\phi_\ell^{(4)}(\textbf{r})$ can be found in
Appendix~\ref{apen:potential}.

\begin{figure*}[!t]
\includegraphics[width=1.0\linewidth]{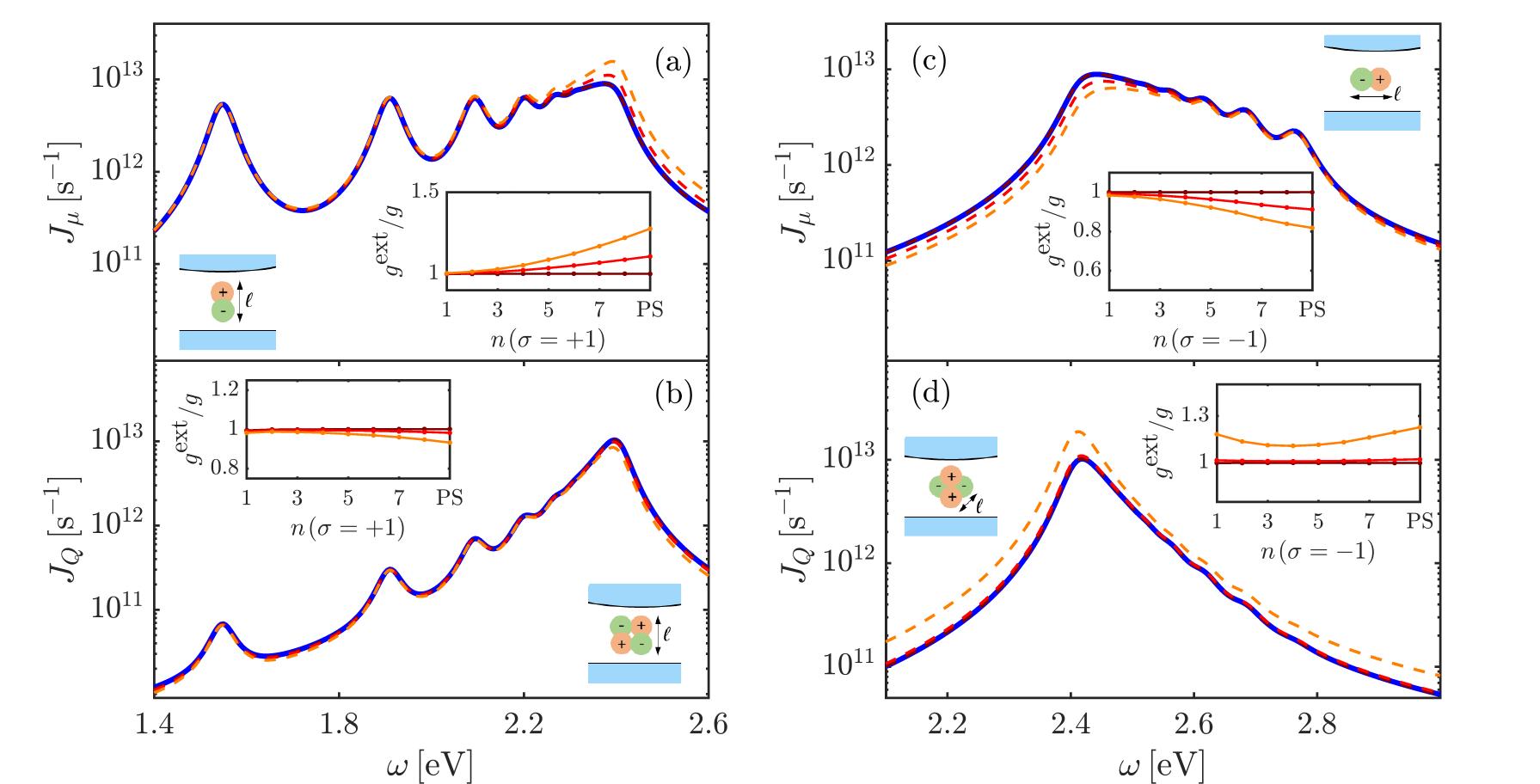}
\vspace{-0.3 cm} \caption{Size effects in the spectral density
(main panels) and coupling strengths (insets) for the same NPoM
cavity and QE parameters as in Fig.~\ref{fig:3}. (a) Dipolar and
(b) quadrupolar QEs with $\alpha = 0$. (c) Dipolar and (d)
quadrupolar QEs with $\alpha=\pi/2$ and $\alpha=\pi/4$,
respectively. The exciton charge distributions are sketched in all
panels. The point-source approximation (blue) and finite-size
calculations for three different $\ell$ are shown: 0.05 nm
(brown), 0.4 nm (red) and 0.6 nm (orange).} \label{fig:7}
\end{figure*}

Figure~\ref{fig:7} reveals the complex phenomenology behind
mesoscopic effects in QE-SP coupling, which depends very much on
the emitter orientation (see sketches in all panels). Dipolar QEs
with $\alpha=0$ (a) and $\alpha=\pi/2$ (c) are displayed in the
top panels, whereas quadrupolar excitons with $\alpha=0$ (b) and
$\alpha=\pi/4$ (d) are shown in the bottom panels. The geometric
and material parameters are the same as in Fig.~\ref{fig:3} (note
that $z_E=0.5\delta$). Spectral densities calculated using the
point-source approximation (blue) are compared against finite-size
charge distributions for different $\ell$: 0.05 nm (brown), 0.4 nm
(red) and 0.6 nm (orange). As expected, the former coincides with
the point-source spectra in all cases, which proves the validity
of Eqs.~\eqref{eq:Pmuext}~and~\eqref{eq:PQext} in the limit
$\ell\rightarrow0$.

Spectral densities for vertical and horizontal dipoles in
Fig.~\ref{fig:7}(a) and (c) show the opposite dependence on
$\ell$, whereas $J_\mu(\omega)$ increases for $\alpha=0$, it
decreases for $\alpha=\pi/2$. These deviations occur mainly at the
pseudomode position, $\omega_{\rm PS}\simeq2.4$ eV, whereas peaks
in $J_\mu(\omega)$ at lower (a) and higher (c) frequencies, which
are associated to low-order SPs with $\sigma=+1$ and $\sigma=-1$,
respectively, are rather insensitive to $\ell$. This is evident in
the insets of both panels, which plot the coupling strengths
obtained from extended calculations normalized to the point-dipole
predictions. Note that they are computed following the same
procedure as descried in Sec.~\ref{sec:density}. We can observe
that $g^{\rm ext}/g\simeq1$ for $n<6$, whereas the ratio increases
(a) or decreases (c) significantly with $\ell$ for larger $n$. The
contrast between both descriptions is maximum at the pseudomode,
which allows us to gain insight into our findings through the map
in Fig.~\ref{fig:6}(c) (evaluated for $\alpha=0$). Indeed, we can
infer that the coupling enhancement in Fig.~\ref{fig:7}(a) is due
to the fact that the exciton charges approach the metal boundaries
as $\ell$ increases for $\alpha=0$, where $g^{PS}_\mu$ is maximum.
On the contrary, they displace laterally, away from the gap center
and towards regions of lower $g^{PS}_\mu$ for $\alpha=\pi/2$,
yielding the coupling reduction in Fig.~\ref{fig:7}(b).

The bottom panels of Fig.~\ref{fig:7} show that, for QEs located
at the gap center ($z_E=0.5\delta$), the impact of finite-size
effects are smaller for quadrupolar excitons than for dipolar
ones. Fig.~\ref{fig:6}(d) shows that $g^{PS}_Q$ is more localized
than its dipolar counterpart at the metal boundaries, which
explains the insensitivity of both $J_Q(\omega)$ and $g^{\rm
ext}/g\simeq1$ to QE dimensions up to $\ell=0.4$ nm for both
orientations. Only for $\ell=0.6$ nm (orange lines) deviations
from the point-quadrupole approximation become apparent, which
again, they take place mainly at the pseudomode frequency, due to
the strongly confined character of the EM fields associated to
high-order SPs. The spectral density and pseudomode coupling are
only slightly lower than the point-quadrupole prediction for
$\alpha=0$, while they are significantly higher for
$\alpha=\pi/4$. This higher impact of mesoscopic effects in
Fig.~\ref{fig:7}(d) can be attributed to two factors. First, the
distance between the nearest point charges in the quadrupole
distribution and the metal boundaries are smaller than in
Fig.~\ref{fig:7}(c). Second, by increasing $\ell$, these charges
(located along the vertical axis) interact more strongly with the
odd ($\sigma=-1$) SPs supported by the cavity, while their
counterparts remain along the $z=0.5\delta$ axis, where $g^{PS}_Q$
is minimum. Let us also stress that the coupling strength
calculations, specially in Fig.~\ref{fig:7}(d), must be taken
carefully. The fact that the pseudomode peak governs completely
the spectral density means that the high-Q resonator
limit~\cite{Waks2010}, inherent to the modal decomposition of
$J_i(\omega)$ in our approach, is not a valid assumption in this
case.

\section{Exciton Population Dynamics} \label{sec:dynamics}

Once we have analyzed the dependence of the spectral density and
coupling strengths on the various parameters of the system, we
explore next the onset of strong coupling between dipolar and
quadrupolar excitons and NPoM cavities. Using our TO approach, we
can parameterize the Hamiltonian governing the coherent QE-SP
interaction
\begin{eqnarray}
\hat{H}_{\rm
sys}&=&\omega_i\hat{\sigma}^{\dag}_i\hat{\sigma_i}+\sum_{n,\sigma}\omega_{n,\sigma}\hat{a}^{\dag}_{n,\sigma}\hat{a}_{n,\sigma}+
\notag
\\
&&+\sum_{n,\sigma}g_{n,\sigma}[\hat{\sigma}^{\dag}_i\hat{a}_{n,\sigma}+\hat{\sigma}_i\hat{a}_{n,\sigma}^{\dag}],\label{eq:Hamiltonian}
\end{eqnarray}
where $\hat{\sigma}_i$ and $\hat{a}_{n,\sigma}$ are the QE
($i=\mu,Q$) and SP annihilation operators (we take $\hbar=1$). The
full density matrix of the system is then given by the master
equation,
\begin{eqnarray}
\frac{\partial \hat{\rho}}{\partial t}=i[\hat{\rho},\hat{H}_{\rm
sys}]+\sum_{n,\sigma}\frac{\gamma_m}{2}\mathcal{L}_{\hat{a}_{n,\sigma}}[\hat{\rho}],\label{eq:mastereq}
\end{eqnarray}
where $\gamma_m$ is the Drude damping rate, and the Lindblad term
$\mathcal{L}_{\hat{a}_{n,\sigma}}[\hat{\rho}]=2\hat{a}_{n,\sigma}\hat{\rho}\hat{a}^{\dag}_{n,\sigma}-\{\hat{a}^{\dag}_{n,\sigma}\hat{a}_{n,\sigma},\hat{\rho}\}$
accounts for the absorption losses experienced by the SP mode with
indices $n$ and $\sigma$. Note that, for the moment, and in order
to gain insight into the plasmon-exciton coupling phenomenology,
we neglect the SP and QE radiative decay. These damping rates are
much smaller than $\gamma_m$, and they do not affect the results
presented below.

We study the temporal evolution of the exciton population,
$n_{E}(t)=\langle
e,\{0\}_{n,\sigma}|\hat{\rho}(t)|e,\{0\}_{n,\sigma}\rangle$, in a
spontaneous emission configuration. Note that
$|e,\{0\}_{n,\sigma}\rangle$ stands for the product of the QE
excited state and the ground state of all SP modes. Thus, we set
the initial density matrix for the system to
$\hat{\rho}(t=0)=|e,\{0\}_{n,\sigma}\rangle\langle
e,\{0\}_{n,\sigma}|$ and investigate the population dynamics. The
exciton population can be calculated through
Eq.~\eqref{eq:mastereq} or, equivalently, through the
Wigner-Weisskopf equation~\cite{Petruccionebook} (beyond the
Markovian approximation
\cite{Li2016,Li2018,Cuartero-Gonzalez2018}) expressed in terms of
the spectral density $J_i(\omega)$
~\cite{Gonzalez-Tudela2014,Li2016}. We will pay special attention
to the ocurrence of non-monotonic, reversible dynamics in
$n_{E}(t)$, which we can relate to the onset of QE-SP strong
coupling and the formation of PEPs at the single emitter level.

\begin{figure}[!t]
\includegraphics[width=1\linewidth]{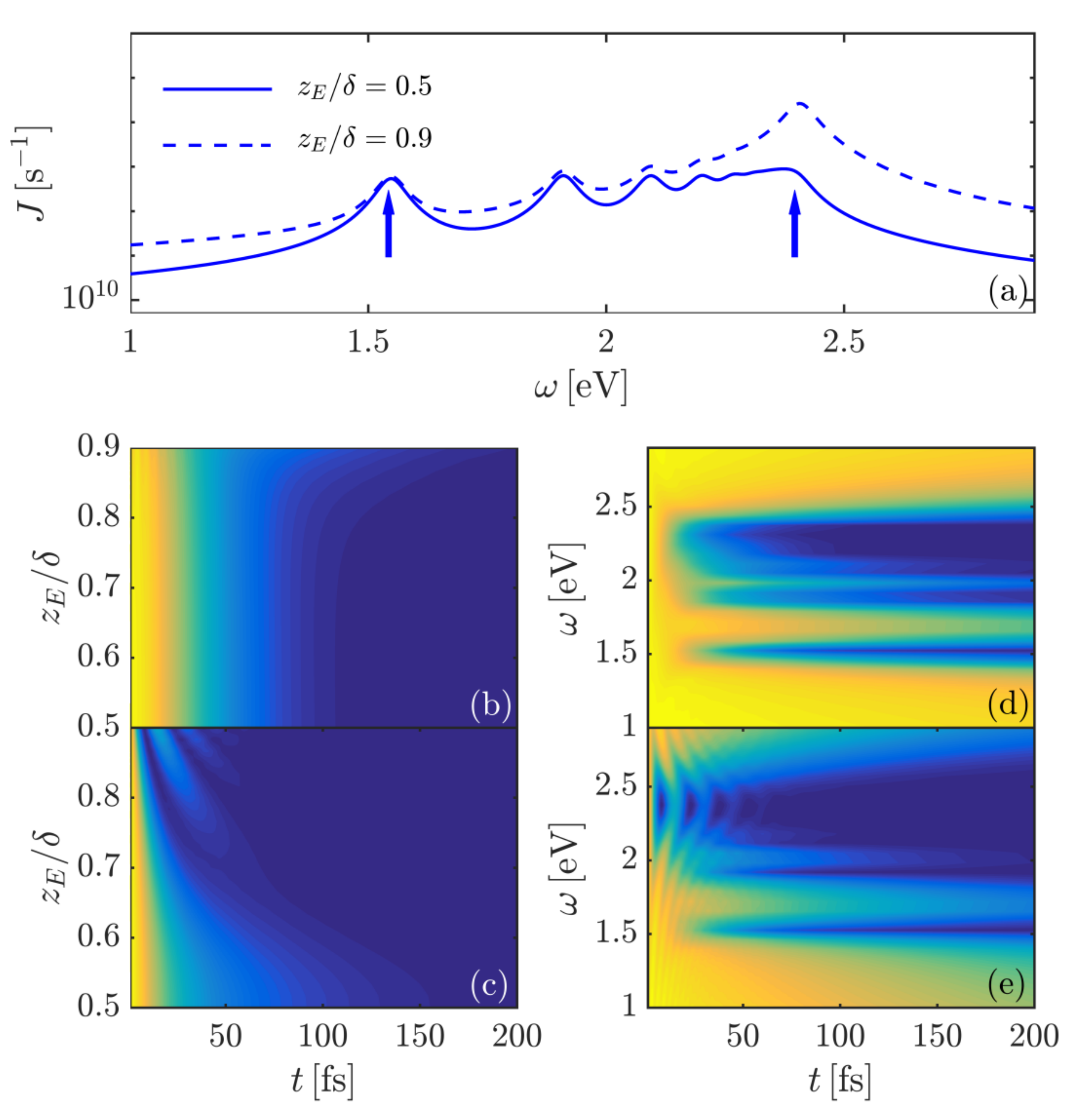}
\vspace{-0.35 cm} \caption{(a) Spectral density for a dipolar QE
at two different positions within the NPoM cavity in
Fig.~\ref{fig:6} ($\mu=0.55$ {e$\cdot$nm}, $\alpha = 0$). QE
population versus time and position for (b)
$\omega_E=\omega_{1,+1}=1.55$ eV and (c) $\omega_E=\omega_{\rm
PS}=2.4$ eV, see vertical arrows in panel (a). QE population
versus time and frequency for (d) $z_E=0.5\delta$ and (e)
$z_E=0.9\delta$. The linear color scale in all contourplots ranges
from $n_E=1$ (yellow) to $n_E=0$ (dark blue).} \label{fig:8}
\end{figure}

Figure~\ref{fig:8} analyzes $n_E(t)$ for a vertically-oriented
dipolar QE ({$\mu=0.55$ e$\cdot$nm}) placed at the gap of a NPoM
cavity with $\delta=0.9$ nm and $D=30$ nm. Fig.~\ref{fig:8}(a)
plots the spectral density at two different positions along the
symmetry axis of the structure, $z_E=0.5\delta$ (solid line) and
$z_E=0.9\delta$ (dashed line). It shows a significant enhancement
in $J_\mu(\omega)$ as the emitter approaches the metal boundaries.
Fig.~\ref{fig:8}(b)~and~(c) display the exciton population as a
function of time and $z_E$ for two different QE frequencies
(indicated by vertical arrows in the top panel):
$\omega_E=\omega_{1,+1}=1.55$ eV and $\omega_E=\omega_{\rm
PS}=2.4$ eV, respectively. If the QE is at resonance with the
lowest-frequency SP (b), $n_E(t)$ undergoes a smooth monotonic
decay. Importantly, this trend barely depends on the QE position.
Taking into account the uniform $g_\mu^{1,+1}$ map in
Fig.~\ref{fig:6}(a), we can conclude that the QE-SP interaction is
governed by this mode in this case. On the contrary, when the QE
is resonant with the plasmonic pseudomode, $n_E$ varies
significantly, see ~\eqref{fig:8}(c). As expected from
$g_\mu^{PS}$ distribution in Fig.~\ref{fig:6}(b), displacing the
QE away from the gap center translates into a faster decay
initially, and in the occurrence of Rabi oscillations in $n_E(t)$
for $z_E>0.7\delta$. Note that their pitch, the Rabi frequency,
diminishes as $z_E$ increases further. They reveal the occurrence
of QE-SP strong-coupling, and the formation of PEPs, the new
eigenstates of the system.

Fig.~\ref{fig:8}(d)~and~(e) explore in a comprehensive manner the
dependence of $n_E(t)$ on the QE natural frequency. The former
corresponds to $z_E=0.5\delta$, and shows that reversible dynamics
does not take place at any $\omega_E$ in this configuration.
Notice though that the decay rate increases abruptly when the
emitter is at resonance with a SP mode. This is particularly
evident at frequencies approaching $\omega_{\rm PS}$. The latter
is evaluated at $z_E=0.9\delta$ and unveils the emergence of
reversible dynamics in the QE population. The Rabi oscillations
become specially apparent in the vicinity of the plasmonic
pseudomode, where the evolution of the QE population within the
first 50 fs exhibits 5 well-defined maxima ($n_E>0.6$) and minima
($n_E\simeq0$). Note that the emergence of these oscillations is
accompanied by an underlying faster decay of the QE population,
which can be linked to strong-coupling version of the phenomenon
of quenching, widely investigated in the weak-coupling
regime~\cite{Delga2014}.

\begin{figure}[!t]
\includegraphics[width=1\linewidth]{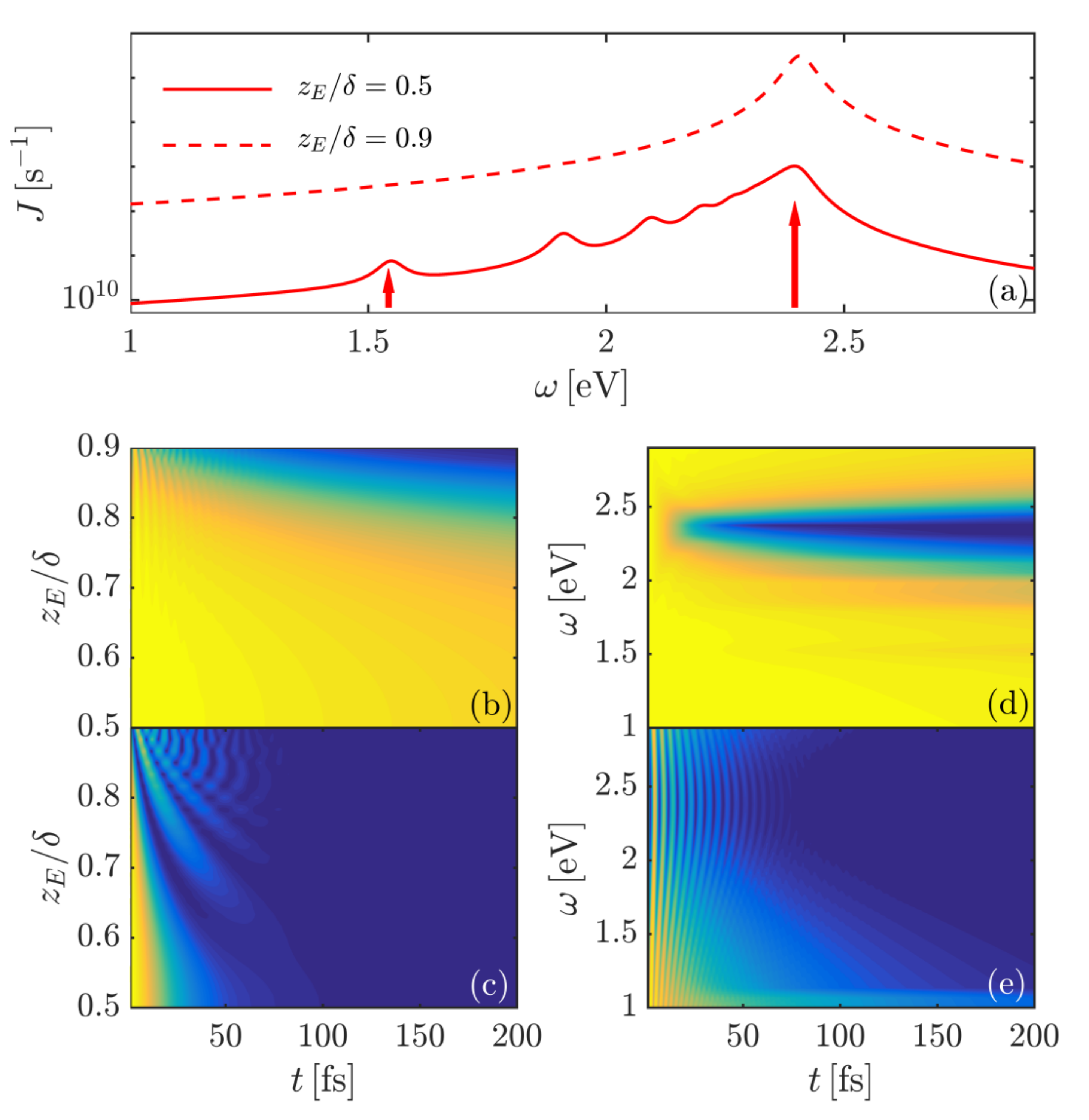}
\vspace{-0.35 cm} \caption{(a) Spectral density for a quadrupolar
QE at two different positions within the NPoM cavity in
Fig.~\ref{fig:3} ($Q=0.75$ {e$\cdot$nm$^2$}$, \alpha = 0$). QE
population versus time and position for (b)
$\omega_E=\omega_{1,+1}=1.55$ eV and (c) $\omega_E=\omega_{\rm
PS}=2.4$ eV, see vertical arrows in panel (a). QE population
versus time and frequency for (d) $z_E=0.5\delta$ and (e)
$z_E=0.9\delta$. The linear color scale in all contourplots ranges
from $n_E=1$ (yellow) to $n_E=0$ (dark blue).} \label{fig:9}
\end{figure}

Figure~\ref{fig:9} reproduces the study in Fig.~\ref{fig:8} but
for quadrupole QEs ({$Q=0.75$ e$\cdot$nm$^2$}, $\alpha=0$).
Fig.~\ref{fig:9}(a) evidences the higher sensitivity of the
quadrupole spectral density on the emitter position. Whereas
several SP maxima are apparent at the gap center (solid line),
$J_Q(\omega)$ is completely governed by the pseudomode at
$z_E=0.9\delta$ (dashed line). Fig.~\ref{fig:9}(b)~and~(c) reveal
that, in agreement with the $g_Q$ contourplots in
Fig.~\ref{fig:6}(c)-(d), $n_E(t)$ depends more strongly on $z_E$
than it does for dipolar QEs. The QE-SP interaction remains in the
weak-coupling regime for $\omega_E=\omega_{1,+1}$ (b), although
the decay rate experiences a strong reduction as $z_E$ increases
(note the small oscillations at large $z_E$, which can be linked
to the spectral detuning between the QE frequency and $\omega_{\rm
PS}$). On the contrary, oscillations in $n_E(t)$ take place when
the QE is only slightly displaced from $z_E=0.5\delta$ for
$\omega_E=\omega_{\rm PS}$ (c). The system enters the
strong-coupling regime in this case, yielding a clear reduction in
the Rabi frequency as the emitter position approaches the metal
surface. Fig.~\ref{fig:9}(d) shows that, regardless of $\omega_E$,
the quadrupolar QE at the gap of the cavity always experience a
monotonic decay (highly Purcell enhanced at the pseudomode). In
contrast, Fig.~\ref{fig:9}(e) proves that in the gap boundaries,
$n_E(t)$ develops Rabi oscillations for all QE frequencies. As
expected, their pitch depends only moderately on $\omega_E$, as
the QE-SP interaction is fully determined by the plasmonic
pseudomode.

\section{Scattering Spectrum} \label{sec:spectrum}

After exploring QE-SP strong-coupling through the temporal
evolution of the exciton population, we turn our attention into
the emergence of PEP signatures in far-field magnitudes, which are
accessible experimentally. Specifically, we model a dark-field
spectroscopy setup~\cite{Lei2012,Ojambati2019}, in which the system is pumped
coherently by a laser field with amplitude $E_L$ polarized along
$z$-direction and with frequency $\omega_L$. The Hamiltonian
describing such experimental configuration is
\begin{eqnarray}
\hat{H}_{\rm exp}=\hat{H}_{\rm sys}+E_{\rm L}e^{-i\omega_{\rm
L}t}\hat{M}^\dagger+E_{\rm L}e^{i\omega_{\rm L}t}\hat{M}
\label{eq:Hexp}
\end{eqnarray}
where $\hat{H}_{\rm sys}$ is given by Eq.~\eqref{eq:Hamiltonian}.
The dipole moment operator of the QE-SP system is
$\hat{M}=\sum_{n}\mu_n\hat{a}_{n,+1} + \mu \hat{\sigma}_\mu$, for
dipolar QEs, and $\hat{M}=\sum_{n}\mu_n\hat{a}_{n,1}$, for
quadrupolar ones. Note that only even SPs ($\sigma=+1$) contribute
to the dipole moment of the NPoM cavity, and that we neglect the
laser excitation of the quadrupole QE. Importantly, due to the
pumping terms, we have $\hat{H}_{\rm exp}=\hat{H}_{\rm exp}(t)$.
This temporal dependence can be removed under an unitary
transformation (see for example Ref.~\cite{Saez-Blazquez2018} for
more details), obtaining $\hat{H'}_{\rm exp}\neq\hat{H'}_{\rm
exp}(t)$ in the laser rotating frame.

In order to compute the dark-field scattering signal, we account
for the radiative losses associated to both SPs and QEs in the
master equation describing the dark-field setup,
\begin{eqnarray}
\frac{\partial \hat{\rho}'}{\partial
t}=i[\hat{\rho}',\hat{H}'_{\rm
exp}]+\sum_{n,\sigma}\frac{\gamma_{n,\sigma}}{2}\mathcal{L}_{\hat{a}_{n,\sigma}}[\hat{\rho}']+\frac{\gamma_i^{\rm
r}}{2}\mathcal{L}_{\hat{\sigma}_i}[\hat{\rho}'], \label{eq:meqexp}
\end{eqnarray}
where $\hat{\rho'}$ is the density matrix in the rotating frame.
Note that Eq.~\eqref{eq:Hexp} incorporates the radiative decay of
both SPs and QE, which were absent in Eq.~\eqref{eq:Hamiltonian}.
The QE radiative decay rates above account for the effect of the
embedding dielectric, having $\gamma_i^{\rm
r}=\sqrt{\epsilon_d}\gamma_i$ ($i=\mu,Q$), where $\gamma_i$ are
the decay rates in vacuum (see Sec.~\ref{sec:density}). The SPs
decay rate have a non-radiative and a radiative component,
$\gamma_{n,\sigma}=\gamma_m+\gamma^{\rm r}_{n,\sigma}$. The latter
is computed by introducing radiative corrections in our TO
approach. Using a procedure very similar to the one presented in
Ref.~\cite{Demetriadou2017} for our NPoM geometry, we obtain
\begin{eqnarray}
\gamma_{n,+1}^{\rm r} &=&  \frac{n \pi D^2 \omega_{n,+1}}{c^2}
\Big(\rho + \sqrt{\rho(\rho + 1)}\Big)^2 \times \notag
\\
&&\times\frac{\omega_p^2 - \omega_{n,1}^2 (\epsilon_{\infty} -
\epsilon_d)}{(\epsilon_{\infty} - \epsilon_d) - (\epsilon_{\infty}
+ \epsilon_d)(\sqrt{\rho} + \sqrt{1+\rho})^{2n}}
\end{eqnarray}
and $\gamma^{\rm r}_{n,-1}=0$. Once $\gamma_{n,+1}^{\rm r}$ are
known, the even SP dipole moments can be obtained by means of the
method of images~\cite{Aubry2011},
having~\cite{Cuartero-Gonzalez2018}
\begin{equation}
\mu_n={\rm
Re}\left\{\frac{\epsilon_m(\omega_{n,+1})+\epsilon_d}{2\epsilon_m(\omega_{n,1})}\right\}
\sqrt{\frac{3\pi\epsilon_0\hbar\gamma^{\rm
r}_{n,+1}c^3}{\sqrt{\epsilon_d}\omega_{n,+1}^3}}.\label{eq:mu}
\end{equation}

Once Eqs.~\eqref{eq:Hexp}~and~\eqref{eq:meqexp} are fully
parameterized by means of TO analytical calculations, we can
obtain the steady-state density matrix of the system by solving
$\partial \hat{\rho}'_{\rm SS}/\partial t=0$. Subsequently, we can
compute the scattering cross section through the square of the
expectation value of the total QE-SP dipole moment operator,
\begin{equation}
\sigma_{\rm sca}(\omega_L)=\langle \hat {M}\rangle_{\rm SS}^2={\rm
Tr}\{\hat{\rho}'_{\rm SS}(\omega_L)\hat{M}\}^2.
\label{eq:sigmascat}
\end{equation}
In the Supplemental Material of Ref.\cite{Cuartero-Gonzalez2018},
we show that Eq.~\eqref{eq:sigmascat}, restricted to the first
excitation manifold and in the limit of low pumping
($E_L\rightarrow0$), reproduces $\sigma_{\rm sca}$ for bare NPoM
cavities.

After a brief description of our calculation of far-field spectra,
we investigate next the scattering properties of the QE-SP hybrid
systems considered in Sec.~\ref{sec:dynamics}. Mimicking the
experimental configuration, we fix the QE frequency at resonance
with the lowest, brightest SP mode, for which $\omega_{1,+1}=1.55$
eV and $\mu_1=46$ e$\cdot$nm, and focus in a narrow spectral
window around it. Note that the exciton population dynamics reveal
that the strongest QE-SP coupling takes place at the plasmonic
pseudomode, but the dark and absorptive character of high-order
SPs hampers its probing through the far-field spectrum around
$\omega_L=\omega_{\rm PS}$. We anticipate though that, despite the
significant detuning, strong-coupling signatures to high-order SPs
can be recognized in $\sigma_{\rm sca}(\omega_L)$ at lower
frequencies.

\begin{figure}[!t]
\includegraphics[width=1\linewidth]{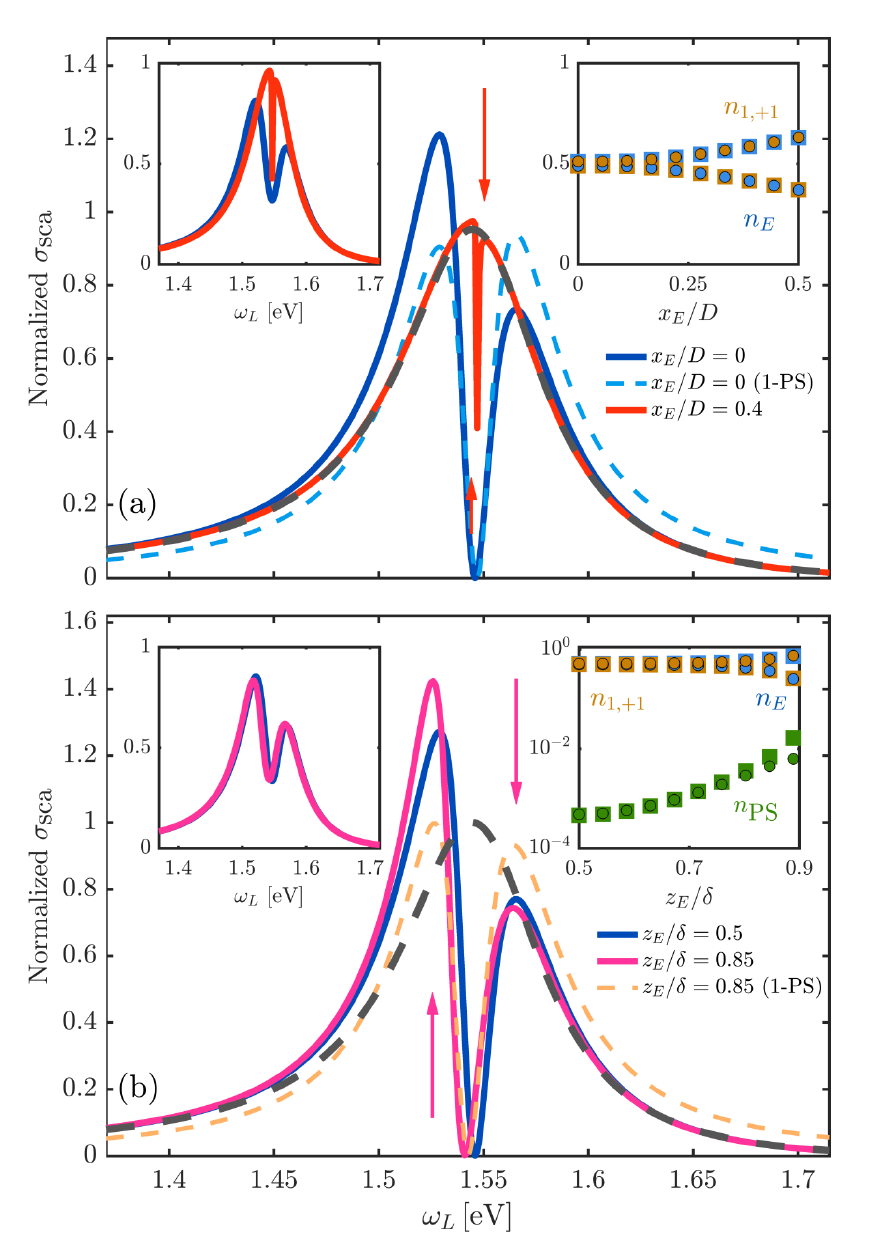}
\vspace{-0.45 cm} \caption{Far-field scattering spectra for a
dipolar QE placed at a NPoM cavity. All parameters are the same as
in Fig.~\ref{fig:8} and $\omega_E=\omega_{1,+1}$. The QE is
displaced away from the gap center along (a) $x$ and (b)
$z$-directions. In both panels, grey dashed and blue solid lines
render the bare cavity cross section and the spectrum for the QE
at the gap center, respectively. Red vertical arrows indicate the
PEP frequencies for the spectrum in red line. Right insets: Square
of the Hopfield coefficients, $n_E$, $n_{1,+1}$ and $n_{\rm PS}$,
for the lower (squares) and upper (circles) PEP as a function of
the QE position. Left insets: $\sigma_{\rm sca}(\omega_L)$ for the
same system configurations as those rendered by solid lines in the
main panels but evaluated at $\gamma_\mu=20$ meV.} \label{fig:10}
\end{figure}

Figure~\ref{fig:10} shows normalized scattering spectra for a
vertically-oriented dipolar QE. The normalization of the cross
section is set so that $\sigma_{\rm sca}(\omega_{1,+1})=1$ for the
bare NPoM structure (in absence of QEs). Grey dashed lines
correspond to the bare cavity, and blue solid ones plot
$\sigma_{\rm sca}(\omega_L)$ when the QE is at the gap center (the
spectra are the same in both panels). The former present a
symmetric maximum centered at the SP frequency. The latter exhibit
a well-defined Rabi doublet structure, with a central minimum at
$\omega_L=\omega_{1,+1}$ and two maxima at the upper (U) and lower
(L) PEP
frequencies~\cite{Savasta2010,Manjavacas2011,Saez-Blazquez2017}.
This splitting is considered the fingerprint of QE-SP
strong-coupling regime, and has been thoroughly analyzed in recent
experimental reports on gap plasmonic
cavities~\cite{Zengin2015,Chikkaraddy2016,Santhosh2016}.

The gap-center spectrum (blue solid line) in Fig.~\ref{fig:10} is
clearly asymmetric, as the maximum below the SP frequency (LPEP)
is significantly higher than the one above it (UPEP). In
Fig.~\ref{fig:10}(a), we analyze the origin of this asymmetry,
recently reported in different plasmonic
systems~\cite{Saez-Blazquez2018,NeumanOpt2018}. The cyan dashed
line plots $\sigma_{\rm sca}(\omega_L)$ obtained by considering
only the lowest SP and the plasmonic pseudomode in the cavity
field. We can observe that the doublet is symmetric in this case,
which allows us to conclude that the height difference of the
peaks in the full calculation originates from the interaction
between the QEs and even SP modes with azimuthal indices between 2
and $n_{\rm min}$ (see Eq.~\eqref{eq:gPS}). The red solid line in
this panel is evaluated for a QE displaced away from the gap
center by $0.4D=12$ nm along $x$-direction. Fig.~\ref{fig:6}(a)
shows that $g^{1,+1}_\mu$ is much lower in this position. The
spectrum overlaps with the bare cavity, except in the vicinity of
$\omega_{1,+1}$, where it develops a Fano-like
profile,~\cite{Ridolfo2010}, characteristic of the weak, or
possibly the intermediate~\cite{Leng2018}, coupling regime. The
red vertical arrows indicate the PEP frequencies (eigenfrequencies
of the Hamiltonian in Eq.~\eqref{eq:Hamiltonian}) in this
configuration. Note that their separation is of the order of
$\gamma^{\rm r}_\mu\simeq10\,\mu$eV. The sharp dip in the spectrum
is a consequence of the weak, coherent interaction between QE and
cavity~\cite{Ridolfo2010}. The right inset plots the square of the
Hopfield coefficients for the LPEP (squares) and the UPEP
(circles) as a function of $x_E/D$. These give the PEP content on
the dipolar QE, $n_E=\langle e,\{0\}_{n,\sigma}|\hat{\rho}_{\rm
SS}(\omega_{\rm PEP})|e,\{0\}_{n,\sigma}\rangle$, (blue dots) and
lowest SP mode, $n_{1,+1}=\langle g,1_{1,+1}|\hat{\rho}_{\rm
SS}(\omega_{\rm PEP})|g,1_{1,+1}\rangle$ (yellow dots). They show
that, due to the reduction experienced by the QE-SP coupling, the
lower (upper) PEP becomes more QE-like (SP-like) as $x_E/D$
increases.

Figure~\ref{fig:10}(b) explores the effect of moving the dipolar
QE vertically. Red solid line plots $\sigma_{\rm sca}(\omega_L)$
for emitter positions very close to the metal surface
($z_E=0.9\delta$). We can observe that both the Rabi splitting and
the difference between LPEP and UPEP scattering maxima remain very
similar to the ones at the gap center. On the contrary, the whole
doublet structure has shifted significantly to lower frequencies
(note that the scattering minima is no longer at $\omega_{1,+1}$).
Orange dashed line plots the same spectrum but considering only
the lowest SP and the pseudomode in the evaluation of
Eq.~\eqref{eq:sigmascat}. The position of the doublet is the same
as in the full calculation but, once again, the asymmetry in the
peaks height has vanished. This fact agrees with our
interpretation, which links the differences in the scattering
maxima with intermediate ($2\geq n\geq n_{\rm min}$) even SP
modes. This approximate spectrum exhibits the same redshift as the
exact one. Taking Fig.~\ref{fig:6}(a)~and~(c) into account, we can
attribute this shifting of the Rabi doublet to the stronger
coupling between the QE and the plasmonic pseudomode caused by the
vertical displacement. The squared Hopfield coefficients in the
right inset of this panel show that, similarly to
Fig.~\ref{fig:10}(a), the balance between $n_E$ and $n_{1,+1}$ in
both PEPs is lost as $z_E$ increases. Importantly, in contrast to
the lateral displacement, this unbalance is accompanied here by an
exponential growth of $n_{\rm PS}=\langle
g,1_\textrm{PS}|\hat{\rho}_{\rm SS}(\omega_{\rm
PEP})|g,1_\textrm{PS}\rangle$ (green dots). This verifies that,
indeed, the redshift experienced by the scattering features
originates from the stronger interaction between the QE and the
plasmonic pseudomode. In fact, it can be interpreted as a result
of the anticrossing between the UPEP and another, even higher
frequency, PEP (not analyzed here) that is located at $\omega_{\rm
PS}$ in the limit of low QE-SP
coupling~\cite{Cuartero-Gonzalez2018}.

The left insets in Fig.~\ref{fig:10}(a)-(b) plot spectra for the
same system configurations as those rendered in solid lines in the
main panels, but replacing the radiative decay rate
$\gamma_\mu^{\rm r}$ in Eq.\eqref{eq:meqexp}, which is of the
order of $0.5\,\mu$eV (at $\omega_E=2$ eV), by a much larger rate
$\gamma_\mu=20$ meV. This way, we account for both the
non-radiative decay and dephasing experienced by QEs in a very
simplified fashion~\cite{Saez-Blazquez2017}. Note that the
linewidth of J-aggregates, which exhibit sharp absorption and
emission bands, are of the order of 10-30 meV at room
temperature~\cite{Valleau2012}. Therefore, by introducing a more
realistic description of QEs, our approach yields smoother
scattering spectra, which resemble those in recent experimental
reports~\cite{Chikkaraddy2016,Santhosh2016,Roller2016,Gross2018,Park2019}.
Importantly, the physical discussion regarding the fingerprint of
PEP formation in $\sigma_{\rm sca}(\omega_L)$ above remains valid
for these insets.

\begin{figure}[!t]
\includegraphics[width=1\linewidth]{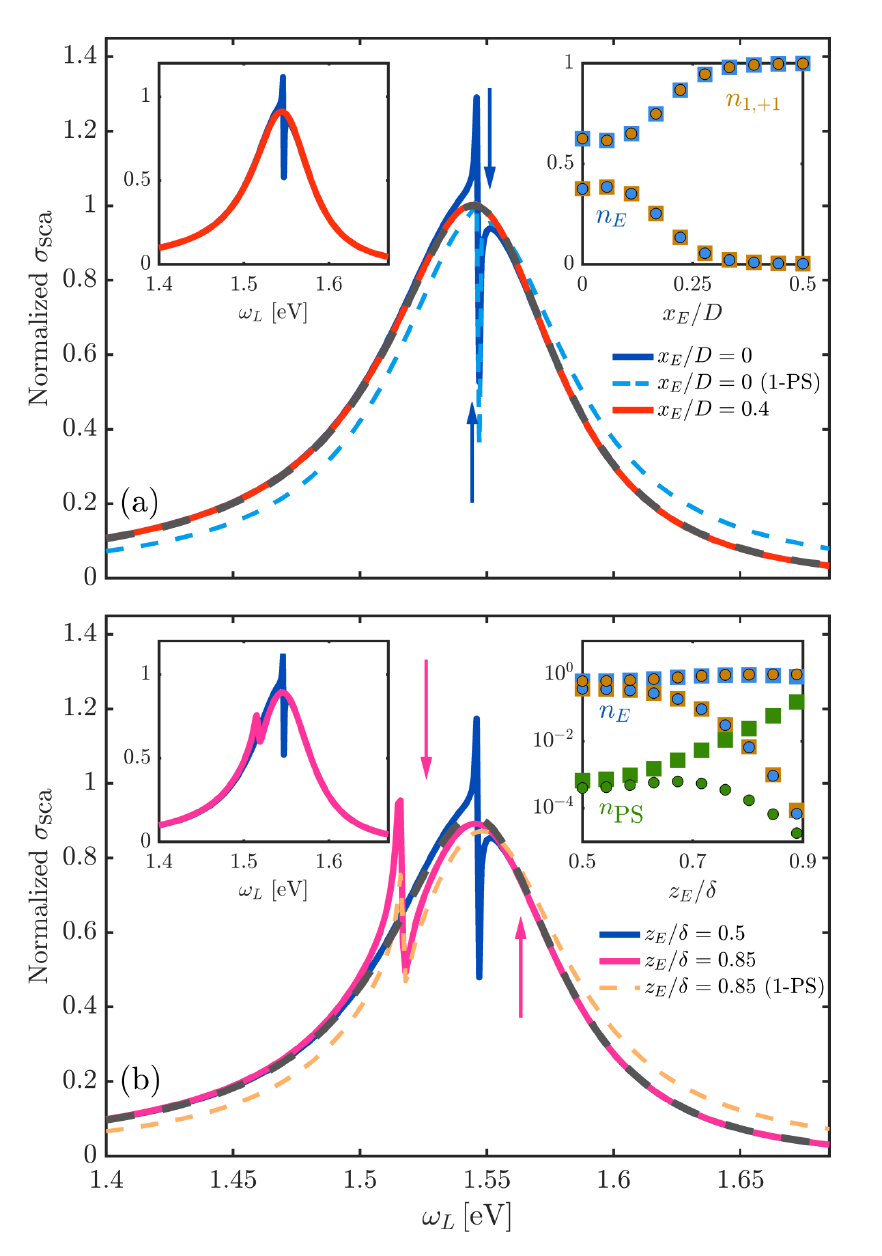}
\vspace{-0.45 cm}\caption{Far-field scattering spectra for a
quadrupolar QE placed at a NPoM cavity. All parameters are the
same as in Fig.~\ref{fig:9} and $\omega_E=\omega_{1,+1}$. The QE
is displaced away from the gap center along (a) $x$ and (b)
$z$-directions. In both panels, grey dashed and blue solid lines
render the bare cavity cross section and the spectrum for the QE
at the gap center, respectively. In the top (bottom) panel, blue
(red) vertical arrows indicate the PEP frequencies at the blue
(red) spectrum. Right insets: Square of the Hopfield coefficients,
$n_E$, $n_{1,+1}$ and $n_{\rm PS}$, for the lower (squares) and
upper (circles) PEP as a function of the QE position. Left insets:
$\sigma_{\rm sca}(\omega_L)$ for the same system configurations as
those rendered by solid lines in the main panels but evaluated at
$\gamma_Q=2$ meV.} \label{fig:11}
\end{figure}

Figure~\ref{fig:11} presents an analysis similar to the one in
Fig.~\ref{fig:10} but for quadrupolar QEs (note the narrower
frequency window). Grey dashed and blue solid lines in both panels
plot the bare cavity cross section and the spectrum for
$z_E=0.5\delta$ (all system parameters are the same as in
Fig.~\ref{fig:9}). In contrast to its dipolar counterpart, the
gap-center spectrum does not exhibit a Rabi doublet, but a sharp
Fano profile at $\omega_L=\omega_{1,+1}$. This is due to the fact
that $g_Q^{1,+1}\ll g_\mu^{1,+1}$ at the gap center, see
Fig.~\ref{fig:6}. As a result of the weaker coupling, the system
eigenfrequencies (see vertical arrows) are close together in this
case. Cyan dashed lines in Fig.~\ref{fig:11}(a) render
$\sigma_{\rm sca}(\omega_L)$ for the same configuration, but
including only the lowest SP and the pseudomode in the
calculation. The deviations from the exact result are apparent
mainly at the scattering minimum, which reveals that intermediate
SPs play a more minor role than in dipolar QEs. Red solid line
plots the $\sigma_{\rm sca}(\omega_L)$ for $x_E=0.4D=12$ nm. As
expected from the tight localization of $g_Q^{1,+1}$ at the NPoM
gap in Fig.~\ref{fig:6}(b), this spectrum coincides with the
scattering cross section of the bare cavity, as QE-SP interactions
vanish in this position. The square of the PEP Hopfield
coefficients in the right inset shows that the system remains in
the weak-coupling regime for all $x_E/D$ values. They demonstrate
that the LPEP (UPEP) collapses rapidly into the quadrupole exciton
(lowest SP mode) as the emitter moves away from the center of the
gap.

The sensitivity of the scattering cross section to variations in
the vertical position of the quadrupolar QE is analyzed in
Fig.~\ref{fig:11}(b). Red solid and orange dashed lines plot
$\sigma_{\rm sca}(\omega_L)$ at $z_E=0.85\delta$ obtained from the
full NPoM plasmonic spectrum and including only the lowest SP and
pseudomode contributions in the calculation, respectively. The
differences between them are even smaller than at the gap center,
see Fig.~\ref{fig:11}(a). We can observe that, by approaching the
emitter to the metal surface, the Fano-like profile in the blue
solid line shifts to lower frequencies, but no Rabi doublet
structure emerges in the spectrum. This means that the interaction
between the QE and the lowest SP remains in the weak-coupling
regime, despite the enhancement experienced by their coupling
strength. Fig.~\ref{fig:6}(b)-(d) reveal that $g_Q^{\rm PS}$ grows
much faster than $g_Q^{1,+1}$ with $z_E$, which explains why the
main effect observed in $\sigma_{\rm sca}(\omega_L)$ is the
red-shift of the Fano feature. Again, this occurs due to the
anticrossing with another PEP, whose initial content is mainly
pseudomode~\cite{Cuartero-Gonzalez2018}. The square of the LPEP
Hopfield coefficients in the right inset shows that for larger
$z_E/\delta$, $n_{1,+1}$ decreases, while $n_{\rm PS}$ increases,
modifying the inherent character of this polariton, which now
emerges from the hybridization of the QE exciton and the plasmonic
pseudomode. On the contrary, the UPEP collapses into the lowest,
bright SP in this process, decoupling completely from the
quadrupole QE.

As in the exploration of dipolar QEs, we have introduced two left
insets in Fig.~\ref{fig:11}(a)-(b), which render the scattering
cross sections for the same system configurations as those plotted
in solid lines in the main panels, but for a non-vanishing
quadrupolar QE decay rate, $\gamma_Q=2$ meV. This allows us to
show how the sharpness of the Fano-like spectral features in the
main panels is reduced once a finite linewidth is introduced in
the QE model.

\section{Conclusions} \label{sec:conclusions}

We have presented a transformation optics approach that exploits
two-dimensional conformal mapping to obtain a full analytical,
insightful description of plasmon-exciton interactions in a
nanoparticle-on-a-mirror cavity. Two different quantum emitters,
supporting only dipolar or only quadrupolar transitions, have been
thoroughly analyzed and compared. We have firstly computed the
nanocavity spectral densities for both emitter families, which can
be decomposed in terms of lorentzian contributions. This enables
us to identify the plasmon-exciton coupling strengths for the full
nanocavity electromagnetic spectrum, which becomes naturally
quantized. Next, we have characterized in detail the dependence of
plasmon-exciton coupling strengths on the emitter position and
orientation. Special attention has been paid to mesoscopic effects
taking place when the dimensions of the exciton charge
distribution are comparable to the gap of the structure. Finally,
the onset of the strong-coupling regime and the formation of
plasmon-exciton polaritons has been investigated in two different,
complementary, studies. First, we have revealed the occurrence of
Rabi oscillations in the temporal evolution of the exciton
population in a spontaneous emission configuration. Second, we
have shown the emergence of a Rabi doublet structure in the
dark-field scattering spectrum of the nanocavity-emitter system
under laser illumination. We believe that our findings can serve
as a guidance for the design and interpretation of experiments
aiming to harness plasmon-exciton strong-coupling phenomena at the
single emitter level.

\section*{Acknowledgement}

This work has been funded by the Spanish MINECO under contracts
FIS2015-64951-R, MDM-2014-0377-16-4 and through the ``Mar\'ia de
Maeztu'' programme for Units of Excellence in R\&D
(MDM-2014-0377), as well as the EU Seventh Framework Programme
under Grant Agreement FP7- PEOPLE-2013-CIG-630996.

\appendix
\section{Quasi-static Potential Calculation} \label{apen:potential}

Here, we provide the analytical expressions behind the Purcell
factor and spectral density calculations in Section~\ref{sec:TO}
and ~\ref{sec:density}, as well as the scattering potential for a
neutral multiple point-charge distribution introduced in
Section~\ref{sec:finite}.

The source potentials describing the array of transformed
point-sources in Fig.~\ref{fig:1}(b) can be written as

\begin{widetext}
\begin{eqnarray}
&\phi_{\mu}^{\rm S}(\varrho') =&
\frac{1}{2\pi\epsilon_0\epsilon_d} \sum_n \text{Re}\bigg\lbrace
\frac{\mu'_{x'}+i\mu'_{z'}}{(\varrho^{'*} - \varrho^{'*}_{E}+2\pi
n
i)} \bigg\rbrace, \label{eq:transmu}\\
&\phi_Q^{\rm S}(\varrho') =&
\frac{1}{2\pi\epsilon_0\epsilon_d}\sum_n
\text{Re}\bigg\lbrace\frac{Q'_{x'x'}+iQ'_{x'z'}}{(\varrho^{'*} -
\varrho^{'*}_{E}+2\pi n i)^2} \bigg\rbrace, \label{eq:transQ}
\end{eqnarray}
where
\begin{equation}
\varrho'_{E} = \ln\left(\Big|1+\tfrac{{2D\sqrt{\rho(1+\rho)}}(i x_E + (z_E - s))}{x_{E}^2+(z_{E}-s)^2}\Big| \right)+\notag + i\arctan\left({\tfrac{x_{E}}{ \tfrac{x_{E}^2+(z_{E}-s)^2}{{2D\sqrt{\rho(1+\rho)}}} + (z_{E}-s)}}\right),
\end{equation}
and the transformed dipolar and quadrupolar moments have the form
(for small distances)
\begin{eqnarray}
\mu'_{x'}+i\mu'_{z'} &=&\tfrac{\mu_x+i\mu_z}{(\varrho_{E}-is)\Big(\frac{i(\varrho_{E}-is)}{2D\sqrt{\rho(1+\rho)}} - 1\Big)} \\
Q'_{x'x'}+iQ'_{x'z'} &=&
\tfrac{Q_{xx}+iQ_{xz}}{(\varrho_{E}-is)^2\Big(\frac{i(\varrho_{E}-is)}{2D\sqrt{\rho(1+\rho)}} - 1\Big)^2}
\end{eqnarray}
Note that the conformal nature of the mapping in
Eq.~\eqref{eq:map} preserves the character of the original
excitation potential: a single dipole (quadrupole) source
transforms into a periodic array of identical dipole (quadrupole)
sources.\\

In order to solve Laplace's Equation in the transformed frame and
obtain the total quasi-static potentials, we Fourier transform
Eqs.~\eqref{eq:transmu}~and~\eqref{eq:transQ}. Then, we impose
that the scattered potentials have the same spatial dependence as
the propagating SPs sustained by the metal-dielectric-metal
geometry. Applying continuity conditions at the metal-dielectric
interfaces, and performing an inverse Fourier transform, we obtain
the scattered potentials in transformed space. Using $\phi^{\rm
sc}_{\mu,Q}(\varrho,\omega)=\phi'^{\rm
sc}_{\mu,Q}(\varrho'(\varrho),\omega)$, we obtain their analytical
expression in the NPoM frame

\begin{eqnarray}
\phi^{\rm sc}_{\mu}(\varrho,\omega) &=&
\frac{1}{2\pi\epsilon_0\epsilon_d} \bigg(\frac{\epsilon_m(\omega)
- \epsilon_d}{\epsilon_m(\omega) - \epsilon_d} \bigg) \sum_{n =
1}^{\infty} \frac{1}{(\sqrt{\rho} + \sqrt{1+\rho})^{4n} -
\big(\frac{\epsilon_{m}(\omega)-\epsilon_d}{\epsilon_m(\omega)+
\epsilon_d}\big)^2} \times \notag \\ && \times\bigg[
\bigg(\frac{\epsilon_m(\omega) - \epsilon_d}{\epsilon_m(\omega) -
\epsilon_d} \bigg) \textrm{Re} \lbrace
(\mu'_{x'}+i\mu'_{z'})A_n^{-}(\varrho,\varrho_E)\rbrace +
(\sqrt{\rho} + \sqrt{1+\rho})^{2n} \textrm{Re} \lbrace
(\mu_{x'}'^{ *}-i\mu_{z'}'^{ *}) B_n^{-}(\varrho,\varrho_E)
\rbrace \bigg] \label{eq:pot_mu_sc}
\end{eqnarray}
and
\begin{eqnarray}
\phi^{\rm sc}_Q(\varrho,\omega) &=&
\frac{1}{2\pi\epsilon_0\epsilon_d} \bigg(\frac{\epsilon_m(\omega)
- \epsilon_d}{\epsilon_m(\omega) - \epsilon_d} \bigg) \sum_{n =
1}^{\infty} \frac{n}{(\sqrt{\rho} + \sqrt{1+\rho})^{4n} -
\big(\frac{\epsilon_m(\omega)-\epsilon_d}{\epsilon_m(\omega)+
\epsilon_d}\big)^2} \times \notag \\
&& \times \bigg[ \bigg(\frac{\epsilon_m(\omega) -
\epsilon_d}{\epsilon_m(\omega) - \epsilon_d} \bigg)\textrm{Re}
\lbrace (Q'_{x'x'}+iQ'_{x'z'}) A_n^{+}(\varrho,\varrho_e)\rbrace -
(\sqrt{\rho} + \sqrt{1+\rho})^{2n} \textrm{Re} \lbrace
(Q'^{*}_{x'x'}-iQ'^{*}_{x'z'})B_n^{+}(\varrho,\varrho_E)\rbrace
\bigg], \notag \\ \label{eq:pot_Q_sc}
\end{eqnarray}
where
\begin{eqnarray}
A_n^{\pm}(\varrho,\varrho_E) &=& \bigg[\bigg(\frac{(2iD\sqrt{\rho(1+\rho)} + \varrho - is)(\varrho_E - is )}{(2iD\sqrt{\rho(1+\rho)} + \varrho_E - is)(\varrho- is)} \frac{}{} \bigg)^{-n} \pm \bigg(\frac{(2iD\sqrt{\rho(1+\rho)} + \varrho - is)(\varrho_E - is )}{(2iD\sqrt{\rho(1+\rho)} + \varrho_E - is)(\varrho- is)} \frac{}{} \bigg)^{n}  \bigg] \\
B_n^{\pm}(\varrho,\varrho_E) &=& \bigg[ e^{2n\Delta}
\bigg(\frac{(2iD\sqrt{\rho(1+\rho)} +
\varrho-is)(\varrho_{E}-is)}{(2iD\sqrt{\rho(1+\rho)} +
\varrho_{E}-is)(\varrho-is)}\bigg)^{-n} \pm e^{-2n\Delta}
\bigg(\frac{(2iD\sqrt{\rho(1+\rho)} +
\varrho-is)(\varrho_{E}-is)}{(2iD\sqrt{\rho(1+\rho)} +
\varrho_{E}-is)(\varrho-is)}\bigg)^{n} \bigg]\\
\end{eqnarray}
and
\begin{equation}
\Delta = \ln(\sqrt{\rho} + \sqrt{1 + \rho}) - \textrm{Re}
\bigg\lbrace \ln\bigg(
1+\frac{2iD\sqrt{\rho(1+\rho)}}{\varrho_{E}-is} \bigg) \bigg\rbrace
\end{equation}
The total potentials can then be written as $\phi^{\rm
tot}_{i}(\varrho)=\phi_{i}^{\rm S}(\varrho)+\phi^{\rm
sc}_i(\varrho)\simeq\phi^{\rm sc}_i(\varrho)$ with $i=\mu,Q$.

Finally, the scattered potential for a neutral distribution of
$N=2,4$ point charges $q_k$ (with $|q_k|=|q|$) located at
positions $\varrho_k$, separated by distances given by the
displacement $\ell$ reads
\begin{eqnarray}
\phi_\ell^{(N)}(\varrho,\omega) &=&
\frac{|q|}{2\pi\epsilon_0\epsilon_d}
\bigg(\frac{\epsilon_m(\omega) - \epsilon_d}{\epsilon_m(\omega) -
\epsilon_d} \bigg)\sum_{n = 1}^{\infty} \frac{1/n}{(\sqrt{\rho} +
\sqrt{1+\rho})^{4n}-\big(\frac{\epsilon_{m}(\omega)-\epsilon_d}{\epsilon_{m}(\omega)+\epsilon_d}\big)^2}\times \notag \\
&&\times\bigg[\bigg(\frac{\epsilon_m(\omega) -
\epsilon_d}{\epsilon_m(\omega) - \epsilon_d} \bigg)
\textrm{Re}\lbrace A^0_n(\varrho) \rbrace - (\sqrt{\rho} +
\sqrt{1+\rho})^{2n}\textrm{Re} \lbrace B^0_n(\varrho)
\rbrace\bigg] \label{eq:pot_ext}
\end{eqnarray}
with
\begin{equation}
A_n^{0}(\varrho)=\sum_{k=1}^{N}{\rm
sign}(q_k)A_n^{+}(\varrho,\varrho_k)\quad {\rm and}\quad
B_n^{0}(\varrho)=\sum_{k=1}^{N}{\rm
sign}(q_k)B_n^{+}(\varrho,\varrho_k)
\end{equation}

\section{Expressions for light-matter coupling strengths} \label{apen:gs}

By reshaping the quasi-static potentials in
Eqs.~\eqref{eq:pot_mu_sc},~\eqref{eq:pot_Q_sc} and
~\eqref{eq:pot_ext}, we obtain the following analytical
expressions for the light-matter coupling strengths that weight
the various SP contributions to the spectral density in
Equation~\eqref{eq:lorentzians}
\begin{eqnarray}
g_{\mu}^{n,\sigma} &=& \sqrt{ \frac{4\sigma n D^2\rho(1+\rho)\mu^2
\omega^2_{n,\sigma}}{3\pi^2 \hbar c \epsilon_0}\frac{1 +
\xi_{n,\sigma}}{\epsilon_{\infty} + \epsilon_d \xi_{n,\sigma}}
\frac{\textrm{Re}\lbrace \textrm{K}_n(\alpha)\rbrace + \sigma
\textrm{Re}
\lbrace\Lambda_n(\alpha)\rbrace}{(\sqrt{\rho} + \sqrt{1 + \rho})^{2n} - \sigma}}, \notag \\
\\
g_{Q}^{n,\sigma} &=& \sqrt{\frac{8\sigma n^3 D^4 (\rho(1+\rho))^2
Q^2  \omega^2_{n,\sigma}}{45 \pi^2 \hbar c \epsilon_0}\frac{1 +
\xi_{n,\sigma}}{\epsilon_{\infty} + \epsilon_d \xi_{n,\sigma}}
\frac{\textrm{Re} \lbrace \textrm{K}'_n(\alpha) \rbrace + \sigma
\textrm{Re} \lbrace \Lambda'_n(\alpha) \rbrace}{(\sqrt{\rho} +
\sqrt{1 + \rho})^{2n} - \sigma}}
\notag \\
\end{eqnarray}
where $\alpha$ is the angle defining the QE orientation and
\begin{eqnarray}
\textrm{K}_n(\alpha) &=& - \frac{(\sin{\alpha} + i\cos{\alpha})^2}{(\varrho_{E}-is)^2
    (2iD\sqrt{\rho(1+\rho)} + \varrho_{E}-is)^2}   \\
\Lambda_n(\alpha) &=& \frac{ \cosh (2n\Delta)}{|\varrho_{E}-is|^2
|2iD\sqrt{\rho(1+\rho)} + \varrho_{E}-is|^2} \\
\textrm{K}^{'}_n(\alpha) &=& - \frac{
(\sin{2\alpha}+i\cos{2\alpha})^2}{(\varrho_{E}-is)^4
\big(2iD\sqrt{\rho(1+\rho)} + \varrho_{E}-is\big)^4} \\
\Lambda^{'}_n(\alpha) &=& \frac{\cosh(2n\Delta)
    + \frac{2D\sqrt{\rho(1+\rho)} - 2(i\varrho_{E} + s)}{2D\sqrt{\rho(1+\rho)} n} \sinh(2n\Delta)
    }
{|\varrho_{E}-is|^4|2iD\sqrt{\rho(1+\rho)} + \varrho_{E}-is|^4}
\end{eqnarray}
\end{widetext}

\end{document}